\documentclass{article}
\usepackage{amssymb}

\usepackage{graphicx}
\usepackage{amsmath}

\newtheorem{theorem}{Theorem}
\newtheorem{acknowledgement}[theorem]{Acknowledgement}

\newtheorem{conjecture}[theorem]{Conjecture}
\newtheorem{corollary}[theorem]{Corollary}

\newtheorem{definition}[theorem]{Definition}

\newtheorem{lemma}[theorem]{Lemma}

\newtheorem{proposition}[theorem]{Proposition}
\newtheorem{remark}[theorem]{Remark}

\newenvironment{proof}[1][Proof]{\textbf{#1.} }{\ \rule{0.5em}{0.5em}}
\input{tcilatex}

\begin{document}

\title{Quantum\ Feedback Networks: \\
Hamiltonian Formulation }
\author{J. Gough$^{1}$, M.R. James$^{2}$}
\date{}
\maketitle

\begin{abstract}
A quantum network is an open system consisting of several component
Markovian input-output subsystems interconnected by boson field channels
carrying quantum stochastic signals. Generalizing the work of Chebotarev\
and Gregoratti, we formulate the model description by prescribing a
candidate Hamiltonian for the network including details the component
systems, the field channels, their interconnections, interactions and any
time delays arising from the geometry of the network. (We show that the
candidate is a symmetric operator and proceed modulo the proof of
self-adjointness.) The model is non-Markovian for finite time delays, but in
the limit where these delays vanish we recover a Markov model and thereby
deduce the rules for introducing feedback into arbitrary quantum networks.
The type of feedback considered includes that mediated by the use of beam
splitters. We are therefore able to give a system-theoretic approach to
introducing connections between quantum mechanical state-based input-output
systems, and give a unifying treatment using non-commutative fractional
linear, or M\"{o}bius, transformations.
\end{abstract}

1) Institute for Mathematical and Physical Sciences, University of Wales,
Aberystwyth, Ceredigion, SY23 3BZ, Wales

2) Department of Engineering, Australian National University, Canberra, ACT
0200, Australia

\section{Introduction}

The aim of this paper is to contribute to our understanding of quantum
feedforward and feedback networks by introducing algebraic rules describing
how to obtain an effective model for a network starting from the canonical
description of the component devices as unconnected systems and the
prescription of the interconnections between these systems. In the open
systems approach to quantum mechanics, a unitary dynamics is given for a
quantum mechanical system (e.g. atom, optical cavity, quantum dot, etc.) and
its environment (e.g. optical field). When the auto-correlation time of the
environment processes is small we can employ quantum stochastic
approximations and work with a quantum stochastic unitary evolution \cite{HP}
with associated Heisenberg-Langevin equations of motion \cite{GQCLT}. The
appropriate way to think of the open system is as an input-output system 
\cite{Gardiner Collett} where the input process is a causal field
representing the environment and the output is the scattered field after
interaction with the system. Cascading such systems is a basic example of
feedforward \cite{Carmichael}\cite{Gardiner} \ More generally we can
consider feedback connections \cite{Wiseman Feedback}\cite{Belavkin CCR}\cite
{Bouten vHandel James}\cite{GBS}\cite{GJ}\cite{Warszawski Wiseman Mabuchi} 
\cite{S Lloyd}.

The natural generalization of this is to consider a graph with quantum
fields propagating along the edges and quantum mechanical systems at the
vertices. The simplest quantum network is a single system with input and
output, and for Markov models the evolution is described in terms of a
Hudson-Parthasarathy quantum stochastic unitary adapted process \cite{HP}
with the system space as initial space and the inputs as noise. There is an
alternative description based on the Chebotarev-Gregoratti Hamiltonian model 
\cite{Cheb98}\cite{Gregoratti}. The simplest nontrivial quantum network will
then be two components cascaded in series, and for finite time delays this
will be non-Markovian: there are several technical and conceptual
difficulties in modelling this in terms of standard (quantum) stochastic
calculus. To resolve these issues, we extend the Chebotarev-Gregoratti
Hamiltonian model to networks, incorporating the various interconnections
and time delays into the boundary conditions that define the domain of the
Hamiltonian. It is the authors' opinion that this offers the only feasible
way to address the topologically non-trivial situation of quantum feedback
induced using beam-splitter devices. We will show that the limit of small
time delays leads to a Chebotarev-Gregoratti Hamiltonian of the type
associated with Markovian models. From this we deduce the rules for
introducing feedback/feedforward connections into assemblies of Markovian
components starting from the component model description. The natural
mathematical language to describe this is in terms of non-commutative
fractional linear transformations of the type introduced by C.L. Siegel.

The notion of quantum feedback for quantum input-output systems has been
around in one form or another since the late 1970's and has had a major
influence on theoretical physics and engineering considerations relating to
the rapidly developing field of quantum feedback control. Our results give a
system-theoretic approach to introducing feedback. An important step towards
a general theory of feedforward and feedback connections was made in the
papers of Gardiner \cite{Gardiner} and Carmichael \cite{Carmichael} who
considered quantum optical networks consisting of cascade-connected
components with no gauge couplings, and Yanagisawa and Kimura \cite{YK1}, 
\cite{YK2} who studied the situation where the plants are multi-dimensional
oscillator systems and the external inputs are Bose fields coupling to the
plants via emission/absorption interactions. Yanagisawa and Kimura were able
to exploit the linearity of the dynamics and apply transfer function
techniques to the resulting networks. The present paper deals with general
quantum dynamical systems with gauge couplings and is not restricted to
linear systems. The field channels are assumed to carry quantum stochastic
signals that satisfy the canonical It\={o} table.

In section 2 we review the quantum stochastic and Hamiltonian models for
open quantum systems, and provide some structural results concerning the
parameters used to define the models. Our general Hamiltonian description of
quantum networks is given in section 3. In section 4 we show how edges can
be eliminated to provide simpler Markovian network models. Key examples are
given in section 5, and concrete topological rules are inferred in section 6.

\section{Quantum Markov Input-Output Components}

The concept of quantum input-output systems originates from two independent
sources: the quantum theory of filtering where the output field is the
object of indirect nondemolition measurement \cite{Belavkin79}\cite
{Belavkin83}, and the theory of quantum amplifiers \cite{Caves}. The latter
theory was generalized to a quantum network by Yurke and Denker \cite{Yurke
Denker} where the network is second quantized quantum wire model with quanta
satisfying a Klein-Gordon equation with Kirchhoff boundary conditions at the
vertices (though there is no quantum system associated with the vertices!).
Similarly, starting from a Lagrangian formulation, Gardiner and Collett, cf. 
\cite{Gardiner Collett}, developed the theory of quantum electromagnetic
input-output fields interacting with a\ quantum mechanical system at the
origin. We may consider this set up as the simplest network consisting of an
input channel and an output channel meeting at a vertex (the system). In the
dipole approximation they derive a Langevin equations for the canonical
observables of the system.\ Following a rotating wave approximation, and a
low frequency limit they obtain a quantum white noise theory suitable for
quantum optics models and which is formally equivalent to a
Hudson-Parthasarathy quantum stochastic evolution for a quantum diffusion 
\cite{HP}. The formulation of the boundary conditions at the vertex is then
a crucial aspect of the model prescription and relates the output field to
the input and the system degrees of freedom.

\subsection{Quantum Stochastic Process Description}

The \textit{system} is modelled as quantum mechanical with Hilbert space $%
\frak{h}$. The inputs and outputs are carried along semi-infinite \textit{%
quantum channels} and are modelled as quantum field signals. Specifically,
we model the field quanta as propagating along the channels with constant
velocity $c$ in the direction specified by the arrows. We may then
parameterize both the input and output lines by a single geometric parameter 
$t$ measuring the \textit{arc-time} taken to reach the system. A single
component is sketched below as a two port device having an input and an
output port.

\begin{center}
\setlength{\unitlength}{.04cm}
\begin{picture}(120,45)
\label{pic1}
\thicklines
\put(45,10){\line(0,1){20}}
\put(45,10){\line(1,0){30}}
\put(75,10){\line(0,1){20}}
\put(45,30){\line(1,0){30}}
\thinlines
\put(48,20){\vector(-1,0){45}}
\put(120,20){\vector(-1,0){20}}
\put(120,20){\line(-1,0){48}}
\put(50,20){\circle{4}}
\put(70,20){\circle{4}}
\put(100,26){input}
\put(48,35){system}
\put(10,26){output}
\end{picture}
%

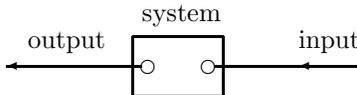
Figure 1: input-output component
\end{center}

The inputs correspond to the half line $\mathbb{R}^{+}=\left( 0,\infty
\right) $ as they have yet to reach the system $\left( t=0\right) $, while
the outputs correspond to $\mathbb{R}^{-}=\left( -\infty ,0\right) $ as they
have already passed through the system. Signals therefore have state space $%
L_{\frak{K}}^{2}\left( \mathbb{R},dt\right) $ where $\frak{K}$ is a fixed
Hilbert space called the \textit{multiplicity space}. We shall generally
consider $\frak{K}=\mathbb{C}^{n}$ which means that we have $n$
distinguishable particles. We shall consider an indefinite number of these
quanta in the wire so that the Hilbert space is the Fock space 
\begin{equation*}
\frak{F}=\Gamma \left( L_{\frak{K}}^{2}\left( \mathbb{R},dt\right) \right) ,
\end{equation*}
where $\Gamma \left( \cdot \right) $ is the bosonic Fock space functor. Note
that Fock spaces have the functorial property 
\begin{equation*}
\Gamma \left( \frak{h}_{1}\oplus \frak{h}_{2}\right) \cong \Gamma \left( 
\frak{h}_{1}\right) \otimes \Gamma \left( \frak{h}_{2}\right) ,
\end{equation*}
and so $\frak{F}=\frak{F}_{\text{in}}\otimes \frak{F}_{\text{out}}$ where $%
\frak{F}_{\text{in}}=\Gamma \left( L_{\frak{K}}^{2}\left( \mathbb{R}%
^{+},dt\right) \right) $ and $\frak{F}_{\text{out}}=\Gamma \left( L_{\frak{K}%
}^{2}\left( \mathbb{R}^{-},dt\right) \right) $.

\subsection{Process Description of a Single Markov Component}

Let us fix the multiplicity number as $n$ and take $\left\{ e_{j}:j=1,\cdots
,n\right\} $ as a basis for $\frak{K}=\mathbb{C}^{n}$. We denote by $%
A_{i}\left( t\right) \triangleq A\left( e_{i}\otimes 1_{\left[ 0,t\right]
}\right) $ and $A_{i}^{\dag }\left( t\right) \triangleq A^{\dag }\left(
e_{i}\otimes 1_{\left[ 0,t\right] }\right) $ the operators describing the 
\textit{annihilation} and \textit{creation} of a quantum in the $i$th
channel over the time interval $\left[ 0,t\right] $, respectively. The
operator describing the scattering from the $j$th channel\ to the $i$th
channel over the time interval $\left[ 0,t\right] $ is denoted by $\Lambda
_{ij}\left( t\right) $. In particular, $N_{i}\left( t\right) =\Lambda
_{ii}\left( t\right) $ is the observable corresponding to the number of
quanta in the $i$th channel over this time.

We now consider a \textit{quantum stochastic evolution} as a unitary adapted
process $\left\{ V\left( t\right) :t\geq 0\right\} $ on $\frak{h}\otimes 
\frak{F}$ in the sense of Hudson and Parthasarathy \cite{HP} arising as the
solution to the quantum stochastic differential equation (QSDE) 
\begin{equation*}
dV=\left( dG\right) V,\quad V\left( 0\right) =1,
\end{equation*}
where 
\begin{eqnarray}
dG\left( t\right) &=&\left( \mathsf{S}_{ij}-\delta _{ij}\right) \otimes
d\Lambda _{ij}\left( t\right) +\mathsf{L}_{i}\otimes dA_{i}^{\dag }\left(
t\right)  \notag \\
&&-\mathsf{L}_{i}^{\dag }\mathsf{S}_{ij}\otimes dA_{j}\left( t\right) -(%
\frac{1}{2}\mathsf{L}_{i}^{\dag }\mathsf{L}_{i}+i\mathsf{H})\otimes dt
\end{eqnarray}
With \textsf{$S$}$=\mathsf{S}_{ij}\otimes |e_{i}\rangle \langle e_{j}|$
unitary on $B\left( \frak{h}\otimes \frak{K}\right) $, \textsf{$L$}$=\mathsf{%
L}_{i}\otimes \langle e_{i}|\in B\left( \frak{h}\otimes \frak{K,h}\right) $%
,and $\mathsf{H}\in B\left( \frak{h}\right) $ self-adjoint. It is convenient
to write these as matrices 
\begin{equation*}
\mathsf{S}\triangleq \left( 
\begin{array}{ccc}
\mathsf{S}_{11} & \dots & \mathsf{S}_{1n} \\ 
\vdots &  & \vdots \\ 
\mathsf{S}_{n1} & \cdots & \mathsf{S}_{nn}
\end{array}
\right) ,\mathsf{L}\triangleq \left( 
\begin{array}{c}
\mathsf{L}_{1} \\ 
\vdots \\ 
\mathsf{L}_{n}
\end{array}
\right) .
\end{equation*}
We introduce the processes 
\begin{eqnarray*}
j_{t}\left( \mathsf{X}\right) &\triangleq &V^{\dag }\left( t\right) \left[ 
\mathsf{X}\otimes 1\right] V\left( t\right) , \\
\tilde{A}_{i}\left( t\right) &\triangleq &V^{\dag }\left( t\right) \left[
1\otimes A_{i}\left( t\right) \right] V\left( t\right) .
\end{eqnarray*}
for $\mathsf{X}\in B\left( \frak{h}\right) $. They satisfy the QSDEs 
\begin{eqnarray}
dj_{t}\left( \mathsf{X}\right) &=&j_{t}\left( \mathcal{L}\left( \mathsf{X}%
\right) \right) dt+j_{t}(\mathsf{S}_{ji}^{\dag }\left[ \mathsf{X},\mathsf{L}%
_{j}\right] )dA_{i}^{\dag }+j_{t}\left( [\mathsf{L}_{i}^{\dag },\mathsf{X}]%
\mathsf{S}_{ij}\right) dA_{j}  \notag \\
&&+j_{t}\left( \mathsf{S}_{ki}^{\dag }\left( \mathsf{X}-1\right) \mathsf{S}%
_{kj}\right) d\Lambda _{ij}\left( t\right) , \\
d\tilde{A}_{i}\left( t\right) &=&j_{t}\left( \mathsf{S}_{ij}\right)
dA_{j}\left( t\right) +j_{t}\left( \mathsf{L}_{i}\right) dt.
\end{eqnarray}
where $\mathcal{L}\left( \mathsf{X}\right) =\frac{1}{2}\mathsf{L}_{i}^{\dag }%
\left[ \mathsf{X},\mathsf{L}_{i}\right] +\frac{1}{2}\left[ \mathsf{L}%
_{i}^{\dag },\mathsf{X}\right] \mathsf{L}_{i}-i\left[ \mathsf{X},\mathsf{H}%
\right] $.

\subsection{System Parameters}

The triple $\left( \mathsf{S},\mathsf{L},\mathsf{H}\right) $, which
determines the model, is referred to as the set of system parameters. The
coefficients of the QSDE are assembled into\ the following square matrices
of $\left( 1+n\right) $ dimensions having entries that are operators on $%
\frak{h}$: 
\begin{eqnarray}
\mathbf{G} &\triangleq &\left( 
\begin{array}{cc}
-\frac{1}{2}\mathsf{L}^{\dag }\mathsf{L}-i\mathsf{H} & -\mathsf{L}^{\dag }%
\mathsf{S} \\ 
\mathsf{L} & \mathsf{S-1}
\end{array}
\right) ,  \label{G-def} \\
\mathbf{V} &\triangleq &\mathbf{G}+\mathbf{\Pi }=\left( 
\begin{array}{cc}
-\frac{1}{2}\mathsf{L}^{\dag }\mathsf{L}-i\mathsf{H} & -\mathsf{L}^{\dag }%
\mathsf{S} \\ 
\mathsf{L} & \mathsf{S}
\end{array}
\right) ,  \label{V-def} \\
\mathbf{M} &\triangleq &\mathbf{1}+\mathbf{\Pi G}=\mathbf{1-\Pi }+\mathbf{%
\Pi V}=\left( 
\begin{array}{cc}
1 & 0 \\ 
\mathsf{L} & \mathsf{S}
\end{array}
\right) ,  \label{M-def}
\end{eqnarray}
where $\mathbf{\Pi }=\left( 
\begin{array}{cc}
0 & 0 \\ 
0 & \mathsf{1}
\end{array}
\right) $. We refer to $\mathbf{G}$ as the \textit{It\={o} generator matrix}
of the unitary evolution, and $\mathbf{V}$ as the \textit{model matrix}. The
matrix $\mathbf{M}\ $\ is called the \emph{Galilean transformation}
associated with $\mathbf{G}$.

\begin{definition}
\label{dfn:model} Let $\frak{h}$ and $\frak{K}$ be fixed Hilbert spaces. The
classes of It\={o} generator matrices $\frak{G}\left( \frak{h},\frak{K}%
\right) $ and model matrices $\frak{M}\left( \frak{h},\frak{K}\right) $ are
collections of operators\ $\mathbf{G},\mathbf{V}\in B\left( \frak{h}\otimes
\left( \mathbb{C}\oplus \frak{K}\right) \right) $ of the form (\ref{G-def})
and (\ref{V-def}) respectively with respect to the decomposition $\frak{h}%
\otimes \left( \mathbb{C}\oplus \frak{K}\right) =\frak{h}\oplus (\frak{h}%
\otimes \frak{K})$ for $\mathsf{S}\in B\left( \frak{h}\otimes \frak{K}%
\right) $ unitary, $\mathsf{L}\in B\left( \frak{h},\frak{h}\otimes \frak{K}%
\right) $, and $\mathsf{H}\in B\left( \frak{h}\right) $ self-adjoint.
\end{definition}

It is convenient to set $A^{00}=t,A^{i0}=A_{j}^{\dag },A^{0j}=A_{i}$ and $%
A^{ij}=\Lambda _{ij}$ and write $dX=X_{\alpha \beta }dA^{\alpha \beta }$ for
a general stochastic integral. We adopt the convention that repeated Greek
indices are summed over 0,1,$\cdots ,n$. The coefficients $X_{\alpha \beta }$
can be assembled into a matrix $\mathbf{X}$ of adapted entries. We may
compress the quantum It\={o} table down to $dA^{\alpha \beta }dA^{\mu \nu }=%
\hat{\delta}^{\beta \mu }dA^{\alpha \beta }$ where $\hat{\delta}^{\alpha
\beta }$ is the Hudson-Evans delta which equals unity when $\alpha =\beta
\in \left\{ 1,\cdots ,n\right\} $ and vanishes otherwise. Note that $\hat{%
\delta}^{\alpha \beta }$ are just the coefficients of the matrix $\mathbf{%
\Pi }$. Given stochastic integrals $X,Y$ with matrices $\mathbf{X,Y}$
respectively, the coefficients of the product $XY$ then form the matrix $%
\mathbf{X}Y+X\mathbf{Y}+\mathbf{X}\Pi \mathbf{Y}$.

We may write the generator of the stochastic evolution as $dG\left( t\right)
=\mathsf{G}_{\alpha \beta }dA^{\alpha \beta }\left( t\right) $ and the
isometry and co-isometry conditions are 
\begin{equation}
\mathbf{G}+\mathbf{G}^{\dag }+\mathbf{G^{\dag }\Pi G}=0=\mathbf{G}+\mathbf{G}%
^{\dag }+\mathbf{G\Pi G}^{\dag }.  \label{Unitary condition}
\end{equation}
The expression for $\mathbf{G}$ above then gives the general solution to
this equation. The Heisenberg equation for an initial operator $\mathsf{X}%
\in B\left( \frak{h}\right) $ is then 
\begin{equation*}
j_{t}\left( \mathsf{X}\right) =\mathsf{X}\otimes 1+\int_{0}^{t}j_{s}\left( 
\mathcal{L}_{\alpha \beta }\left( \mathsf{X}\right) \right) \otimes
dA^{\alpha \beta }\left( s\right)
\end{equation*}
and $j_{t}\left( \mathcal{L}_{\alpha \beta }\left( X\right) \right) $ are
the components of a matrix $j_{t}\left( X\mathbf{G}+\mathbf{G}^{\dag }X+%
\mathbf{G}^{\dag }\mathbf{\Pi }X\mathbf{\Pi G}\right) $. The super-operators 
$\mathcal{L}_{\alpha \beta }$ are known as the Evans-Hudson maps.

The output processes are then $\tilde{A}^{\alpha \beta }\left( t\right)
:=V\left( t\right) ^{\dag }\Lambda ^{\alpha \beta }\left( t\right) V\left(
t\right) $ and we deduce that 
\begin{equation*}
d\tilde{A}^{\alpha \beta }\left( t\right) \equiv j_{t}\left( \mathsf{M}%
_{\alpha \mu }^{\dag }\mathsf{M}_{\beta \nu }\right) \,dA^{\mu \nu }\left(
t\right) .
\end{equation*}
This invariance of time, $d\tilde{t}=dt$, is the motivation for the term
``Galilean transformation'', and we now explore some of its properties.

\begin{definition}
\label{dfn:gal-group} The \emph{Galilean group} $\frak{Gal}\left( \frak{h},%
\frak{K}\right) $ is the group of operators of the form (\ref{M-def}) in $%
B\left( \frak{h}\otimes \left( \mathbb{C}\oplus \frak{K}\right) \right) $
where $\mathsf{S}\in B\left( \frak{h}\otimes \frak{K}\right) $ unitary, $%
\mathsf{L}\in B\left( \frak{h},\frak{h}\otimes \frak{K}\right) $.
\end{definition}

The group identity is $\mathbf{I}$ and we readily observe the group laws 
\begin{eqnarray*}
\left( 
\begin{array}{cc}
1 & 0 \\ 
\mathsf{L}_{1} & \mathsf{S}_{1}
\end{array}
\right) \left( 
\begin{tabular}{ll}
$1$ & $0$ \\ 
$\mathsf{L}_{2}$ & $\mathsf{S}_{2}$%
\end{tabular}
\right) &=&\left( 
\begin{tabular}{ll}
$1$ & $0$ \\ 
$\mathsf{L}_{1}+\mathsf{S}_{1}\mathsf{L}_{2}$ & $\mathsf{S}_{1}\mathsf{S}%
_{2} $%
\end{tabular}
\right) , \\
\left( 
\begin{tabular}{ll}
$1$ & $0$ \\ 
$\mathsf{L}$ & $\mathsf{S}$%
\end{tabular}
\right) ^{-1} &=&\left( 
\begin{tabular}{ll}
$1$ & $0$ \\ 
$-\mathsf{S}^{\dag }\mathsf{L}$ & $\mathsf{S}^{\dag }$%
\end{tabular}
\right) .
\end{eqnarray*}

\begin{proposition}
If $\mathbf{M}\in \frak{Gal}\left( \frak{h},\frak{K}\right) $ then we have
the identity $\mathbf{M\Pi M}^{\dag }=\mathbf{\Pi }.$
\end{proposition}

\begin{proposition}
The set $\frak{G}\left( \frak{h},\frak{K}\right) $ is invariant under the
action $\mathbf{G}\mapsto \mathbf{N}^{\dag }\mathbf{GN}$ for all $\mathbf{N}%
\in \frak{Gal}\left( \frak{h},\frak{K}\right) $.
\end{proposition}

\begin{proof}
Let $\mathbf{G}\in \frak{G}\left( \frak{h},\frak{K}\right) $ and $\mathbf{N}%
\in \frak{Gal}\left( \frak{h},\frak{K}\right) $. Setting $\mathbf{G}^{\prime
}=\mathbf{N}^{\dag }\mathbf{GN}$ we see that 
\begin{eqnarray*}
\mathbf{G}^{\prime }+\mathbf{G}^{\prime \dag } &=&\mathbf{N}^{\dag }\left( 
\mathbf{G}+\mathbf{G}^{\dag }\right) \mathbf{N}=-\mathbf{N}^{\dag }\mathbf{%
G\Pi G}^{\dag }\mathbf{N} \\
&\equiv &-\mathbf{N}^{\dag }\mathbf{GN\Pi N}^{\dag }\mathbf{G}^{\dag }%
\mathbf{N=-G}^{\prime }\mathbf{\Pi G}^{\prime \dag }
\end{eqnarray*}
since $\mathbf{N\Pi N}^{\dag }=\mathbf{\Pi }$. Similarly, $\mathbf{G}%
^{\prime }+\mathbf{G}^{\prime \dag }=\mathbf{-\mathbf{G}^{\prime \dag }\Pi G}%
^{\prime }$.
\end{proof}

\begin{proposition}
Let $\mathbf{G}\in \frak{G}\left( \frak{h},\frak{K}\right) $ with associated
Galilean transformation $\mathbf{M}$ $\mathbf{=1}+\mathbf{\Pi G}$. Then 
\begin{equation*}
\mathbf{M}^{\dag }\mathbf{GM=G}.
\end{equation*}
\end{proposition}

\begin{proof}
This follows from the observation 
\begin{equation*}
\left( \mathbf{I}+\mathbf{G}^{\dag }\mathbf{\Pi }\right) \mathbf{G}\left( 
\mathbf{\mathbf{1}+\mathbf{\Pi G}}\right) \mathbf{=G}+\left( \mathbf{G}%
^{\dag }+\mathbf{G}+\mathbf{G}^{\dag }\mathbf{\Pi G}\right) \mathbf{\Pi G}.
\end{equation*}
\end{proof}

\subsection{Hamiltonian Description of a Single Markov Component}

Let us consider the strongly continuous one-parameter unitary group $%
U_{0}\left( t\right) $ performing the time shifts. For instance, taking $%
\varepsilon \left( f\right) $ to be the exponential vector with test
function $f$, we have the action 
\begin{equation*}
U_{0}\left( t\right) u\otimes \varepsilon \left( f\right) =u\otimes
\varepsilon \left( \vartheta _{-t}f\right)
\end{equation*}
where $\vartheta _{t}f\left( \cdot \right) \triangleq f\left( \cdot
-t\right) $. The family $(V\left( t\right) :t\geq 0\}$ the forms a right
unitary cocycle with respect to $U_{0}$, that is $V\left( t+s\right)
=U_{0}\left( -s\right) V\left( t\right) U_{0}\left( s\right) V\left(
s\right) $ $\left( t,s\geq 0\right) $, and we obtain a strongly continuous
unitary group $U$ by setting 
\begin{equation*}
U\left( t\right) =\left\{ 
\begin{array}{cc}
U_{0}\left( t\right) V\left( t\right) , & t\geq 0, \\ 
V^{\dag }\left( -t\right) U_{0}\left( t\right) , & t<0.
\end{array}
\right.
\end{equation*}
The generators of $U$ and $U_{0}$ will be denoted as $H$ and $H_{0}$
respectively. The problem of characterizing $H$ for the class of Hudson and
Parthasarathy quantum stochastic evolutions has been carried out only in
relatively recent times \cite{Cheb98}\cite{Gregoratti} and we now recall its
explicit construction.

A vector $\Phi \in \frak{H}$ will be a sequence $\left( \Phi _{m}\right)
_{m=0}^{\infty }$ where $\Phi _{m}=\Phi _{m}\left( t_{1},\cdots
,t_{m}\right) $ is a $\frak{h}\otimes \frak{K}^{m}$-valued function
completely symmetric under interchange of its arguments and such that 
\begin{equation*}
\sum_{m}\frac{1}{m!}\int_{\mathbb{R}^{m}}\left\| \Phi _{m}\left(
t_{1},\cdots ,t_{m}\right) \right\| _{\frak{h}\otimes \frak{K}%
^{m}}^{2}dt_{1}\cdots dt_{m}<\infty \text{.}
\end{equation*}
Let $\mathbb{R}_{\ast }=\mathbb{R}\backslash \left\{ 0\right\} =\left(
-\infty ,0\right) \cup \left( 0,\infty \right) $ and define the domain $%
W\left( \mathbb{R}_{\ast },\frak{K},\frak{h}\right) $ consisting of vectors
such that each $\Phi _{m}$ is differentiable in each of its arguments and
that

\begin{enumerate}
\item[i)]  $\sum_{j=1}^{m}\frac{\partial }{\partial t_{j}}\Phi _{m}\in \frak{%
h}\otimes L_{\frak{K}}^{2}\left( \mathbb{R}_{\ast }^{m}\right) $, for each $%
m,$

\item[ii)]  $\sum_{m}\frac{1}{m!}\int_{\mathbb{R}_{\ast }^{m}}\left\|
\sum_{j=1}^{m}\frac{\partial }{\partial t_{j}}\Phi _{m}\left( t_{1},\cdots
,t_{m}\right) \right\| _{\frak{h}\otimes \frak{K}^{m}}^{2}dt_{1}\cdots
dt_{m}<\infty $,

\item[iii)]  $\lim_{t_{m+1}\rightarrow 0^{\pm }}\sum_{m}\frac{1}{m!}\int_{%
\mathbb{R}_{\ast }^{m}}\left\| \Phi _{m}\left( t_{1},\cdots
,t_{m},t_{m+1}\right) \right\| _{\frak{h}\otimes \frak{K}^{m+1}}^{2}$ exists.
\end{enumerate}

\noindent (Note that the left and right limits above need not coincide!)

The following operators $H_{0},a\left( t\right) $ for $t\neq 0$, and $%
a\left( 0^{\pm }\right) $, are introduced on the domain $W\left( \mathbb{R}%
_{\ast },\frak{K},\frak{h}\right) $ 
\begin{eqnarray*}
\left( H_{0}\Phi \right) _{m} &=&i\sum_{j=1}^{m}\frac{\partial }{\partial
t_{j}}\Phi _{m}, \\
\left( a_{i}\left( t\right) \Phi \right) _{m}\left( t_{1},\cdots
,t_{m}\right) &=&\left. e_{i}\lrcorner \Phi _{m+1}\left( t_{1},\cdots
,t_{m+1}\right) \right| _{t=t_{m+1}},
\end{eqnarray*}
where $e_{i}\lrcorner $ is the trace operation from $\frak{h}\otimes \frak{K}%
^{m+1}$ down to $\frak{h}\otimes \frak{K}^{m}$ with respect to $e_{i}\in 
\frak{K}$, see \cite{Gregoratti} for more details. The operator $H_{0}$ on
the Sobolev-Fock domain $W\left( \mathbb{R}_{\ast },\frak{K},\frak{h}\right) 
$\ coincides with the generator $H_{0}$ of translation by time shift on the
dense subset for which the right and left hand limits in iii) agree. More
generally we have the relation, which is a consequence of integration by
parts with a jump discontinuity at the origin, for $\Phi ,\Psi \in W\left( 
\mathbb{R}_{\ast },\frak{K},\frak{h}\right) $%
\begin{equation*}
\langle \Phi |H_{0}\Psi \rangle =\langle H_{0}\Phi |\Psi \rangle
+i\sum_{j=1}^{n}\left\langle a_{j}\left( 0^{-}\right) \Phi |a_{j}\left(
0^{-}\right) \Psi \right\rangle -i\sum_{j=1}^{n}\left\langle a_{j}\left(
0^{+}\right) \Phi |a_{j}\left( 0^{+}\right) \Psi \right\rangle
\end{equation*}
with the sum over an arbitrary orthonormal basis $\left\{ e_{i}\right\} $
for the multiplicity space $\frak{K}$. Note that we may formally write \cite
{Cheb98} 
\begin{equation}
H_{0}\equiv \sum_{j=1}^{n}\int_{\mathbb{R}_{\ast }}a_{j}^{\dag }\left(
t\right) i\frac{\partial }{\partial t}a_{j}\left( t\right) .  \label{ChebOp}
\end{equation}

Next, fix a subset $D_{b.c.}\left( \mathsf{S},\mathsf{L}\right) $ of $%
W\left( \mathbb{R}_{\ast },\frak{K},\frak{h}\right) $ consisting of those
vectors $\Phi $ satisfying the boundary conditions 
\begin{equation}
a_{j}\left( 0^{-}\right) \Phi =\mathsf{S}_{jk}a_{k}\left( 0^{+}\right) \Phi +%
\mathsf{L}_{j}\Phi .
\end{equation}
The boundary condition can be written as a Galilean transformation 
\begin{equation*}
a_{\alpha }\left( 0^{-}\right) \Phi =\mathsf{M}_{\alpha \beta }a_{\beta
}\left( 0^{+}\right) \Phi
\end{equation*}
where we include the time case $a_{0}=1$.

\begin{theorem}[Gregoratti]
\cite{Gregoratti} The Hamiltonian $K$ associated with the quantum stochastic
unitary process having the parameters $\left( \mathsf{S},\mathsf{L},\mathsf{H%
}\right) $ has $dom\left( H\right) \cap W\left( \mathbb{R}_{\ast },\frak{K},%
\frak{h}\right) =D_{b.c.}\left( \mathsf{S},\mathsf{L}\right) $ and here it
is given by 
\begin{equation*}
H\Phi =\left( H_{0}+\mathsf{H}-\frac{i}{2}\mathsf{L}_{j}^{\dag }\mathsf{L}%
_{j}-i\mathsf{L}_{j}^{\dag }\mathsf{S}_{jk}a_{k}\left( 0^{+}\right) \right)
\Phi .
\end{equation*}
The Hamiltonian $H$ is essentially self-adjoint on this domain.
\end{theorem}

For our purposes, it is most convenient to write the equation for $H$ and
the boundary condition in terms of the model matrix $\mathbf{V}$ as 
\begin{eqnarray}
-iH\Phi &=&\left( \mathsf{V}_{00}+\mathsf{V}_{0k}a_{k}\left( 0^{+}\right)
-iH_{0}\right) \Phi \equiv \left( \mathsf{V}_{0\beta }a_{\beta }\left(
0^{+}\right) -iH_{0}\right) \Phi ,  \notag \\
a_{j}\left( 0^{-}\right) \Phi &=&\mathsf{V}_{j0}\Phi +\mathsf{V}_{jk}a\left(
0^{+}\right) \Phi \equiv \mathsf{V}_{j\beta }a_{\beta }\left( 0^{+}\right)
\Phi .  \label{Greg}
\end{eqnarray}

\section{Quantum Networks}

The situation sketched in figure 1 will be our simplest example of a quantum
network: a single component system with one input channel terminating at an
input port and one output channel starting at an output. A general quantum
network will consist of several such components connected together and will
typically have time delays, feedforward and feedback connections. The
description will exhibit separate algebraic, topological and geometric
content. While the abstract definition is rather involved, figure 2 below
gives an example of the class of configurations that we wish to consider

\begin{center}
%
\setlength{\unitlength}{.1cm}
\begin{picture}(120,45)
\label{pic2}

\put(22,0){\dashbox(65,45)}
\thicklines
\put(60,30){\line(0,1){10}}
\put(60,30){\line(1,0){20}}
\put(80,30){\line(0,1){10}}
\put(60,40){\line(1,0){20}}
\put(60,5){\line(0,1){20}}
\put(60,5){\line(1,0){20}}
\put(80,5){\line(0,1){20}}
\put(60,25){\line(1,0){20}}
\put(30,5){\line(0,1){15}}
\put(30,5){\line(1,0){20}}
\put(50,5){\line(0,1){15}}
\put(30,20){\line(1,0){20}}
\thinlines
\put(32,10){\vector(-1,0){15}}
\put(62,35){\vector(-1,0){45}}
\put(25,15){\line(1,0){7}}
\put(25,15){\line(0,1){10}}
\put(25,25){\line(1,0){30}}
\put(25,25){\vector(1,0){15}}
\put(55,15){\line(0,1){10}}
\put(55,15){\line(-1,0){7}}
\put(62,10){\line(-1,0){14}}
\put(60,10){\vector(-1,0){5}}
\put(100,10){\line(-1,0){22}}
\put(100,10){\vector(-1,0){10}}
\put(100,20){\line(-1,0){22}}
\put(100,20){\vector(-1,0){10}}
\put(100,35){\line(-1,0){22}}
\put(100,35){\vector(-1,0){10}}
\put(33,10){\circle{2}}
\put(33,15){\circle{2}}
\put(47,10){\circle{2}}
\put(47,15){\circle{2}}
\put(63,10){\circle{2}}
\put(63,35){\circle{2}}
\put(77,10){\circle{2}}
\put(77,20){\circle{2}}
\put(77,35){\circle{2}}
\put(26,12){$s_2$}
\put(26,17){$s_1$}
\put(51,12){$r_2$}
\put(51,17){$r_1$}
\put(56,12){$s_4$}
\put(55,37){$s_3$}
\put(81,12){$r_5$}
\put(81,22){$r_4$}
\put(81,37){$r_3$}
\put(38,12){$C_1$}
\put(67,14){$C_3$}
\put(67,35){$C_2$}
\end{picture}%
%

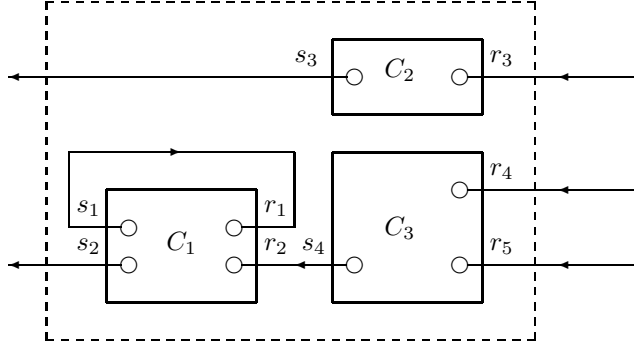
figure 2: A quantum network
\end{center}

We first list the basic features of a network. The network consists of a
collection of components $\mathcal{C}$ that are interconnected in a manner
to be described shortly. Every component $C$ will have at least one input
port and one output port. Let us write $\mathcal{P}_{\text{in}}\left(
C\right) $ and $\mathcal{P}_{\text{out}}\left( C\right) $ as the set of
input and output ports, respectively, for component $C$. For each $r\in 
\mathcal{P}_{\text{in}}\left( C\right) $ we have an associated space $\frak{K%
}_{\text{in}}^{r}$ which is the multiplicity space of the incoming channel.
With a similar notation for the output ports, we impose the constraint 
\begin{equation*}
\frak{K}_{C}\triangleq \bigoplus_{r\in \mathcal{P}_{\text{in}}\left(
C\right) }\frak{K}_{\text{in}}^{r}\equiv \bigoplus_{s\in \mathcal{P}_{\text{%
out}}\left( C\right) }\frak{K}_{\mathrm{out}}^{s}
\end{equation*}
which means that the total multiplicity space of all inputs into a component
equals the corresponding output one. The sets of all input and output ports
in the network are then $\mathcal{P}_{\text{in}}=\cup _{C}\mathcal{P}_{\text{%
in}}\left( C\right) $ and $\mathcal{P}_{\text{out}}=\cup _{C}\mathcal{P}_{%
\text{out}}\left( C\right) $ and we also have 
\begin{equation}
\frak{K}_{\mathrm{net}}\triangleq \bigoplus_{r\in \mathcal{P}_{\text{in}}}%
\frak{K}_{\text{in}}^{r}\equiv \bigoplus_{s\in \mathcal{P}_{\text{out}}}%
\frak{K}_{\mathrm{out}}^{s}.  \label{K net}
\end{equation}

Let $n_{\text{in}}\left( C\right) $ and $n_{\text{out}}\left( C\right) $ be
the number of input and output ports respectively in a component and set $n_{%
\text{in}}=\sum_{C}n_{\text{in}}\left( C\right) $, $n_{\text{out}%
}=\sum_{C}n_{\text{out}}\left( C\right) $. Note that $n_{\text{in}}\left(
C\right) $ and $n_{\text{out}}\left( C\right) $ are both non-zero though
they need not coincide -\ though the total input/output multiplicities $%
\left( \dim \frak{K}_{C}\right) $ must be equal! External fields propagate
into the network along input channels terminating at some of the input
ports. Likewise, the output fields propagate along output channels starting
at some of the output ports. Internally, we also have pairs of input and
output ports connected by further channels. In this way every port is
connected to exactly one channel. The set of interconnections is described
by fixing subsets $\mathcal{R}_{\text{in}}\subset \mathcal{P}_{\text{in}}$
and $\mathcal{R}_{\text{out}}\subset \mathcal{P}_{\text{out}}$ of equal size
and a bijection $\sigma :\mathcal{R}_{\text{out}}\mapsto \mathcal{R}_{\text{%
in}}$. The pair $\left( s,\sigma \left( s\right) \right) $ then determines
an internal channel from output port $s$ to input port $r=\sigma \left(
s\right) $.

Topologically we think of channels as edges and will frequently refer to
them as such, and distinguish the input, output and internal edges. Each
internal edge can be written as a pair $e=\left( s,r\right) $ - $s$ is the 
\emph{source} and $r$ is the \emph{range} - and we will have a corresponding
fixed multiplicity space\ $\frak{K}_{e}$ associated with the channel, and
this must agree with both $\frak{K}_{\text{out}}^{s}$ and $\frak{K}_{\text{in%
}}^{r}$ providing an additional constraint. The remaining ports are $%
\mathcal{Q}_{\text{in}}=\mathcal{P}_{\text{in}}\backslash \mathcal{R}_{\text{%
in}}$ and $\mathcal{Q}_{\text{out}}=\mathcal{P}_{\text{out}}\backslash 
\mathcal{R}_{\text{out}}$. For each $r\in \mathcal{Q}_{\text{in}}$we have a
semi-infinite input edge terminating at that input port, and similar for
each $s\in \mathcal{Q}_{\text{in}}$. We note the identity 
\begin{equation*}
\frak{K}_{\mathrm{ext}}\triangleq \bigoplus_{r\in \mathcal{Q}_{\text{in}}}%
\frak{K}_{\text{in}}^{r}\equiv \bigoplus_{s\in \mathcal{Q}_{\text{out}}}%
\frak{K}_{\mathrm{out}}^{s}
\end{equation*}
which implies that the total multiplicity of all external inputs equals that
of all the outputs. The complete set of edges, including both internal and
external, will be denoted as $\mathcal{E}$.

(In the network sketched in figure 2, we have three components $\mathcal{C}%
=\{C_{1},C_{2},C_{3}\}$ with $\mathcal{P}_{\text{in}}\left( C_{1}\right)
=\left\{ r_{1},r_{2}\right\} $, $\mathcal{P}_{\text{out}}\left( C_{1}\right)
=\left\{ s_{1},s_{2}\right\} $, $\mathcal{P}_{\text{in}}\left( C_{2}\right)
=\left\{ r_{3}\right\} $, $\mathcal{P}_{\text{out}}\left( C_{1}\right)
=\left\{ s_{3}\right\} $ and $\mathcal{P}_{\text{in}}\left( C_{3}\right)
=\left\{ r_{4},r_{5}\right\} $, $\mathcal{P}_{\text{out}}\left( C_{3}\right)
=\left\{ s_{4},s_{5}\right\} $. The total network has $\mathcal{P}_{\text{in}%
}=\left\{ r_{1},r_{2},r_{3},r_{4},r_{5}\right\} $, $\mathcal{P}_{\text{out}%
}=\left\{ s_{1},s_{2},s_{3},s_{4}\right\} $. The interconnections are via
internal edges $e_{1}=\left( s_{1},r_{1}\right) $ and $e_{2}=\left(
s_{4},r_{2}\right) $ and so $\mathcal{R}_{\text{in}}=\left\{
r_{1},r_{2}\right\} $, $\mathcal{R}_{\text{out}}=\left\{ s_{1},s_{4}\right\} 
$ with $\sigma \left( s_{1}\right) =r_{1}$ and $\sigma \left( s_{4}\right)
=r_{2}$, and $\mathcal{Q}_{\text{in}}=\left\{ r_{3},r_{4},r_{5}\right\} $, $%
\mathcal{Q}_{\text{out}}=\left\{ s_{2},s_{3}\right\} $.)

In addition to this topological description, we also provide the arc-time
taken to travel along each internal channel: this determines the various
time delays in the network. We geometrize the edges by applying a local
arctime coordinate to each one. Let $e=\left( s,r\right) $ be an internal
edge then we may assign an arctime parameter $t_{e}$ with range $\left(
T_{r},T_{s}\right) $ with $T_{s}-T_{r}$ being the time taken to travel from
output port $s$ to input port $r$. For input channels terminating at $r\in 
\mathcal{Q}_{\text{in}}$ we have an arctime parameter over the semi-infinite
range $\left( T_{r},\infty \right) $ and likewise for output channels
leaving $s\in \mathcal{Q}_{\text{out}}$ we have an arctime parameter over
the semi-infinite range $\left( -\infty ,T_{s}\right) $. The one-particle
Hilbert space for the field quanta in an edge $e\in \mathcal{E}$ of the
network is then 
\begin{equation*}
L_{\frak{K}_{e}}^{2}\left( e\right) \triangleq \left\{ 
\begin{array}{cc}
L_{\frak{K}_{\text{in}}^{r}}^{2}\left( T_{r},\infty \right) , & e\text{ an
incoming edge terminating at }r\in \mathcal{Q}_{\text{in}}, \\ 
L_{\frak{K}_{\text{out}}^{s}}^{2}\left( -\infty ,T_{s}\right) , & e\text{ an
outgoing edge starting at }s\in \mathcal{Q}_{\text{out}}, \\ 
L_{\frak{K}_{e}}^{2}\left( T_{r\left( e\right) },T_{s\left( e\right)
}\right) , & e=\left( s\left( e\right) ,r\left( e\right) \right) \text{ an
internal edge.}
\end{array}
\right.
\end{equation*}
We set 
\begin{equation*}
L^{2}\left( \mathcal{E}\right) \triangleq \bigoplus_{e\in \mathcal{E}}L_{%
\frak{K}_{e}}^{2}\left( e\right)
\end{equation*}
The Fock space over all these spaces will be denoted as $\frak{F}_{\mathcal{E%
}}$ and by the functorial property factors as 
\begin{equation*}
\frak{F}_{\mathcal{E}}\triangleq \Gamma \left( L^{2}\left( \mathcal{E}%
\right) \right) =\dbigotimes_{e\in \mathcal{E}}\frak{F}_{e}.
\end{equation*}
The Hilbert space for the entire network $\mathcal{N}$\ will then take the
form 
\begin{equation*}
\frak{H}_{\mathcal{N}}\triangleq \frak{h}\otimes \frak{F}_{\mathcal{E}},
\end{equation*}
where $\frak{h}$ is the Hilbert space for all the quantum mechanical degrees
of freedom of the network components.

In figure 2, we have only sketched the interactions between components that
are mediated by the channels. Even though it may appear that the network is
disconnected, the components may still be coupled, say by a Hamiltonian
interaction. We have also stopped short of requiring that the space of
components factors as, say, $\frak{h}=\otimes _{C}\frak{h}_{C}$ and it
convenient not to impose this at this stage.

Up to this point, we have described the flow through the channels from
output to input ports. It still remains to describe the trans-component
flow. This involves the boundary conditions relating the inputs to the
outputs at each component. The most convenient way to describe this is
through the notion of the model matrix. With each component $C$ we associate
a model matrix $\mathbf{V}_{C}\in \frak{M}\left( \frak{h},\frak{K}%
_{C}\right) $.

\begin{definition}
Let $\mathbf{V}_{i}\in \frak{M}\left( \frak{h},\frak{K}_{i}\right) $ for $%
i=1,2$, then the concatenation of model matrices is $\mathbf{V}_{1}\mathbf{%
\boxplus V}_{2}\in \frak{M}\left( \frak{h},\frak{K}_{1}\oplus \frak{K}%
_{2}\right) $ defined by 
\begin{eqnarray*}
&&\left( 
\begin{array}{cc}
-\frac{1}{2}\mathsf{L}_{1}^{\dag }\mathsf{L}_{1}-i\mathsf{H}_{1} & -\mathsf{L%
}_{1}^{\dag }\mathsf{S}_{1} \\ 
\mathsf{L}_{1} & \mathsf{S}_{1}
\end{array}
\right) \mathbf{\boxplus }\left( 
\begin{array}{cc}
-\frac{1}{2}\mathsf{L}_{2}^{\dag }\mathsf{L}_{2}-i\mathsf{H}_{2} & -\mathsf{L%
}_{2}^{\dag }\mathsf{S}_{2} \\ 
\mathsf{L}_{2} & \mathsf{S}_{2}
\end{array}
\right) \\
&\triangleq &\left( 
\begin{array}{ccc}
-\frac{1}{2}\mathsf{L}_{1}^{\dag }\mathsf{L}_{1}-\frac{1}{2}\mathsf{L}%
_{2}^{\dag }\mathsf{L}_{2}-i(\mathsf{H}_{1}+\mathsf{H}_{2}) & -\mathsf{L}%
_{1}^{\dag }\mathsf{S}_{1} & -\mathsf{L}_{2}^{\dag }\mathsf{S}_{2} \\ 
\mathsf{L}_{1} & \mathsf{S}_{1} & 0 \\ 
\mathsf{L}_{2} & 0 & \mathsf{S}_{2}
\end{array}
\right) .
\end{eqnarray*}
\end{definition}

Given the set of component model matrices we may then define the network
model matrix to be 
\begin{equation*}
\mathbf{V}=\boxplus _{C\in \mathcal{C}}\mathbf{V}_{C}\in \frak{M}\left( 
\frak{h},\frak{K}_{\mathrm{net}}\right) .
\end{equation*}
The network model matrix $\mathbf{V}$ relates input ports to output ports
and is introduced independently of the interconnections which go from output
ports to input ports. It takes the standard form 
\begin{equation*}
\mathbf{V}=\left( 
\begin{array}{cc}
-\frac{1}{2}\mathsf{L}^{\dag }\mathsf{L}-i\mathsf{H} & -\mathsf{L}^{\dag }%
\mathsf{S} \\ 
\mathsf{L} & \mathsf{S}
\end{array}
\right) 
\end{equation*}
and, explicitly, for $r\in \mathcal{P}_{\text{in}}$ and $s\in \mathcal{P}_{%
\text{out}}$, we have components 
\begin{eqnarray*}
\mathsf{V}_{sr} &\equiv &\mathsf{S}_{sr}:\frak{h}\otimes \frak{K}_{\text{in}%
}^{r}\mapsto \frak{h}\otimes \frak{K}_{\text{out}}^{s}, \\
\mathsf{V}_{s0} &\equiv &\mathsf{L}_{s}:\frak{h}\mapsto \frak{h}\otimes 
\frak{K}_{\text{out}}^{s}, \\
\mathsf{V}_{0r} &\equiv &-\sum_{s\in \mathcal{P}_{\text{out}}}\mathsf{L}%
_{s}^{\dag }\mathsf{S}_{sr}:\frak{h}\otimes \frak{K}_{\text{in}}^{r}\mapsto 
\frak{h,} \\
\mathsf{V}_{00} &\equiv &-\frac{1}{2}\sum_{s\in \mathcal{P}_{\text{out}}}%
\mathsf{L}_{s}^{\dag }\mathsf{L}_{s}-i\mathsf{H}:\frak{h}\otimes \frak{K}_{%
\text{in}}^{r}\mapsto \frak{h.}
\end{eqnarray*}
It is convenient to adopt the following block matrix representation with
respect to these decompositions $\left( \ref{K net}\right) $ of $\frak{K}_{%
\mathrm{net}}$ 
\begin{equation*}
\mathsf{S}\equiv \left( 
\begin{array}{ccc}
\mathsf{S}_{11} & \mathsf{\dots } & \mathsf{S}_{1n_{\mathrm{in}}} \\ 
\mathsf{\vdots } & \ddots  & \mathsf{\vdots } \\ 
\mathsf{S}_{n_{\mathrm{out}}1} & \mathsf{\cdots } & \mathsf{S}_{n_{\mathrm{%
out}}n_{\mathrm{in}}}
\end{array}
\right) ,\quad \mathsf{L}\equiv \left( 
\begin{array}{c}
\mathsf{L}_{1} \\ 
\mathsf{\vdots } \\ 
\mathsf{L}_{n_{\mathrm{out}}}
\end{array}
\right) .
\end{equation*}
We require that the operators $\mathsf{S}_{sr}$ are contractions with $%
\mathsf{S}$ unitary on $\frak{h}\otimes \frak{K}_{\mathrm{total}}$, and so 
\begin{equation*}
\sum_{s\in \mathcal{P}_{\text{out}}}\mathsf{S}_{sr}^{\dag }\mathsf{S}%
_{sr^{\prime }}=\delta _{rr^{\prime }},\sum_{r\in \mathcal{P}_{\text{in}}}%
\mathsf{S}_{sr}\mathsf{S}_{s^{\prime }r}^{\dag }=\delta _{ss^{\prime }}.
\end{equation*}

It is possible to use the definition of concatenation in reverse in order to
analyze a network model matrix into irreducible components. This process is
substantially more complicated as we have to consider all decompositions $%
\left( \ref{K net}\right) $.

\bigskip

A vector $\Phi \in \frak{H}_{\mathcal{N}}$ can be represented as follows:
for each integer $m\geq 0$ we choose $m$ locations on the edges $\mathcal{E}$
with local arctime coordinates $\{t_{1},\cdots ,t_{m}\}$, say with $t_{j}$
on edge $e\left( j\right) $, to get a vector $\Phi _{m}\left( t_{1},\cdots
,t_{m}\right) \in \frak{h}\otimes \left( \otimes _{j=1}^{m}\frak{K}_{e\left(
j\right) }\right) $ which we view as a subset of $\frak{h}\otimes
L^{2}\left( \mathcal{E}^{m}\right) $, where $\mathcal{E}^{m}$ denotes the $m$%
-fold Cartesian product of $\mathcal{E}$. We now generalize the class of
Sobolev-Fock vectors to networks.

\begin{definition}
The class $W\left( \mathcal{N}\right) $ of Sobolev-Fock functions over a
network $\mathcal{N}$ is the set of vectors $\left( \Phi _{m}\right)
_{m=0}^{\infty }$ in $\frak{H}_{\mathcal{N}}$ such that each $\Phi _{m}$ is
differentiable it each of its arguments and that
\end{definition}

\begin{enumerate}
\item[i)]  $\sum_{j=1}^{m}\frac{\partial }{\partial t_{j}}\Phi _{m}\in \frak{%
h}\otimes L^{2}\left( \mathcal{E}^{m}\right) $, \textit{for each} $m,$

\item[ii)]  $\sum_{m}\frac{1}{m!}\left\| \sum_{k=1}^{m}\frac{\partial }{%
\partial t_{k}}\Phi _{m}\right\| _{\frak{h}\otimes L^{2}\left( \mathcal{E}%
^{m}\right) }^{2}<\infty $,

\item[iii)]  $\lim_{t_{m+1}\rightarrow 0^{\pm }}\sum_{m}\frac{1}{m!}\left\|
\Phi _{m+1}\left( \cdot ,\cdots ,\cdot ,t_{m+1}\right) \right\| _{\frak{h}%
\otimes L^{2}\left( \mathcal{E}^{m+1}\right) }^{2}$ \textit{exists.}
\end{enumerate}

For each internal edge $e=\left( r,s\right) $ we introduce the local
annihilator density $a_{e,j}\left( t\right) $ on the domain of Sobolev-Fock
associated with the edge and the $j$th vector of an orthonormal basis $%
\left\{ e_{j}\right\} $ for the multiplicity space $\frak{K}_{e}$.and $%
T_{r}<t<T_{s}$. We also consider the one-sided limits $T_{r}^{+}$ and $%
T_{s}^{-}$ and collect into the column-block vector of annihilators 
\begin{equation*}
a_{e}(t)\triangleq \left( 
\begin{array}{c}
a_{e,1}\left( t\right)  \\ 
\vdots  \\ 
a_{e,n}\left( t\right) 
\end{array}
\right) ,\quad a_{r}(T_{r}^{+})\triangleq a_{e}(T_{r}^{+}),\quad
a_{s}(T_{s}^{-})\triangleq a_{e}(T_{s}^{-}),
\end{equation*}
where $n=\dim \frak{K}_{e}$. We then define the operator for each internal
edge $e=\left( r,s\right) $ 
\begin{equation*}
H_{0}\left( e\right) =\sum_{j}\int_{T_{r}}^{T_{s}}a_{e,j}^{\dag }\left(
t\right) i\frac{\partial }{\partial t}a_{e,j}\left( t\right)
dt=\int_{T_{r}}^{T_{s}}a_{e}^{\dag }\left( t\right) i\frac{\partial }{%
\partial t}a_{e}\left( t\right) dt.
\end{equation*}
For the semi-finite\ external edges we have the corresponding expression
with either upper or lower limit extended to infinity as appropriate. The
total contribution is then $H_{0}=\sum_{e\in \mathcal{E}}H_{0}\left(
e\right) $ and this generalizes the operator $\left( \ref{ChebOp}\right) $.
We will denote the ampliation of these operators up to $\frak{h}\otimes 
\frak{F}$ by the same symbol, and recall our earlier convention that $%
a_{0}\equiv 1$.

\begin{definition}
The Hamiltonian $H_{\mathcal{N}}$ for a quantum network $\mathcal{N}$ with
model matrix $\mathbf{V}$\ is the operator on $\frak{H}_{\mathcal{N}}$ given
by the specification that all Sobolev-Fock vectors $\Phi $ in its domain
satisfy the system of boundary conditions 
\begin{equation*}
a_{s}\left( T_{s}^{-}\right) \Phi =\sum_{r\in \left\{ 0\right\} \cup 
\mathcal{P}_{\text{in}}}\mathsf{V}_{sr}a_{r}\left( T_{r}^{+}\right) \Phi ,
\end{equation*}
for each component $s\in \mathcal{P}_{\text{out}}$, and on such vectors we
have the action 
\begin{equation*}
-iH_{\mathcal{N}}\Phi =\sum_{r\in \left\{ 0\right\} \cup \mathcal{P}_{\text{%
in}}}\mathsf{V}_{0r}a_{r}\left( T_{r}^{+}\right) \Phi -iH_{0}\Phi .
\end{equation*}
\end{definition}

\begin{lemma}
The operator $H_{\mathcal{N}}$ in the definition of a quantum network
Hamiltonian is symmetric on the restricted domain $dom\left( H_{\mathcal{N}%
}\right) \cap W\left( \mathcal{N}\right) $.
\end{lemma}

\begin{proof}
We now have the network integration-by-parts formula 
\begin{equation*}
\langle \Phi |H_{0}\Psi \rangle -\langle H_{0}\Phi |\Psi \rangle
=i\sum_{s\in \mathcal{P}_{\text{out}}}\left\langle a_{s}\left(
T_{s}^{-}\right) \Phi |a_{s}\left( T_{s}{}^{-}\right) \Psi \right\rangle
-i\sum_{r\in \mathcal{P}_{\text{in}}}\left\langle a_{r}\left(
T_{r}^{+}\right) \Phi |a_{r}\left( T_{r}^{+}\right) \Psi \right\rangle
\end{equation*}
for Sobolev-Fock vectors $\Phi ,\Psi $. $H_{\mathcal{N}}$ is symmetric on
the set of Sobolev-Fock vectors satisfying the boundary conditions, indeed, 
\begin{gather*}
\langle \Phi |H_{\mathcal{N}}\Psi \rangle -\langle H_{\mathcal{N}}\Phi |\Psi
\rangle =i\left\langle \Phi |\mathsf{V}_{00}\Psi \right\rangle
-i\left\langle \mathsf{V}_{00}\Phi |\Psi \right\rangle +i\sum_{s\in \mathcal{%
P}_{\text{out}}}\left\langle a_{s}\left( T_{s}^{-}\right) \Phi |a_{s}\left(
T_{s}{}^{-}\right) \Psi \right\rangle \\
+i\sum_{r\in \mathcal{P}_{\text{in}}}\left\{ \left\langle \Phi |\mathsf{V}%
_{0r}a_{r}\left( T_{r}^{+}\right) \Psi \right\rangle -\left\langle \mathsf{V}%
_{0r}a_{r}\left( T_{r}^{+}\right) \Phi |\Psi \right\rangle -\left\langle
a_{r}\left( T_{r}^{+}\right) \Phi |a_{r}\left( T_{r}^{+}\right) \Psi
\right\rangle \right\}
\end{gather*}
and, substituting in for $a_{s}\left( T_{s}{}^{-}\right) \Psi $ and $%
a_{s}\left( T_{s}{}^{-}\right) \Phi $ and using the identities $\mathsf{V}%
_{00}+\mathsf{V}_{00}^{\dag }+\sum_{s\in \mathcal{P}_{\text{out}}}\mathsf{V}%
_{s0}^{\dag }\mathsf{V}_{s0}=0$, $\mathsf{V}_{0r}+\sum_{s\in \mathcal{P}_{%
\text{out}}}\mathsf{V}_{s0}^{\dag }\mathsf{V}_{sr}$, we find the right hand
side vanishes.
\end{proof}

\bigskip

We may rewrite these equations alternatively as 
\begin{eqnarray*}
-iH\Phi  &=&-(\frac{1}{2}\sum_{s\in \mathcal{P}_{\text{out}}}\mathsf{L}%
_{s}^{\dag }\mathsf{L}_{s}+i\mathsf{H}+iH_{0})\Phi -\sum_{s\in \mathcal{P}_{%
\text{out}}}\sum_{r\in \mathcal{P}_{\text{in}}}\mathsf{L}_{s}^{\dag }\mathsf{%
S}_{sr}a_{r}\left( T_{r}^{+}\right) \Phi . \\
a_{s}\left( T_{s}^{-}\right) \Phi  &=&\sum_{r\in \mathcal{P}_{\text{in}}}%
\mathsf{S}_{sr}a_{r}\left( T_{r}^{+}\right) \Phi +\mathsf{L}_{s}\Phi .
\end{eqnarray*}
We wish to identify $H_{\mathcal{N}}$ as the physical Hamiltonian for the
network, that is, show that it is essentially self-adjoint. This is a
reasonable assumption given that it is a direct second quantization of
quantum graph models encountered in the literature. At the moment, we do not
have a proof of this fact but formulate it as a conjecture.

\begin{conjecture}
We now assume that $H_{\mathcal{N}}$ defines an essentially self-adjoint
operator.
\end{conjecture}

The Hamiltonian $H_{\mathcal{N}}$ contains all the physical information
about the network, including the interactions and interconnections.
Generally speaking, the wave operator $V=U_{0}^{\dag }U$ will not determine
a quantum stochastic evolution of Hudson Parthasarathy type as the model is
typically no longer Markovian. We shall show that, in the limit in which
time delays along the internal paths vanish, we recover a Markovian model
which is easily identifiable from $H$ and which will have the
Chebotarev-Gregoratti form.

\section{Eliminating Internal Edges in the Zero Time Delay Limit}

In this section we show how simpler Markovian models can be obtained by
eliminating edges in a zero time delay limit. We achieve this by showing
first how to eliminate one edge, and then showing that all edges may be
eliminated, with the final Markovian model independent of the order in which
the eliminations were performed.

\begin{theorem}
Let $e_{0}=\left( r_{0},s_{0}\right) $ be an internal channel with time
delay $\tau _{0}=T_{s_{0}}-T_{r_{0}}\geq 0$ in a quantum network $\mathcal{N}
$ for which $1-\mathsf{V}_{s_{0}r_{0}}$ is invertible. In the limit $\tau
_{0}\rightarrow 0^{+}$, the network reduces to $\mathcal{N}_{\text{red}}$ in
which the input and output ports are $\mathcal{P}_{\text{in}}\backslash
\left\{ r_{0}\right\} $ and $\mathcal{P}_{\text{out}}\backslash \left\{
s_{0}\right\} $ and the edge $e_{0}$ eliminated. (In the case where $r_{0}$
and $s_{0}$ are initially in different components, then the components
merge.) The reduced model matrix $\mathbf{V}^{\text{red}}$ then has the
components 
\begin{equation}
\mathsf{V}_{\alpha \beta }^{\text{red}}=\mathsf{V}_{\alpha \beta }+\mathsf{V}%
_{\alpha r_{0}}\left( 1-\mathsf{V}_{s_{0}r_{0}}\right) ^{-1}\mathsf{V}%
_{s_{0}\beta },  \label{Vred}
\end{equation}
for $\beta \in \left\{ 0\right\} \cup \mathcal{P}_{\text{in}}\backslash
\left\{ r_{0}\right\} $ and $\alpha \in \left\{ 0\right\} \cup \mathcal{P}_{%
\text{out}}\backslash \left\{ s_{0}\right\} $.
\end{theorem}

\begin{proof}
For $\Phi $ in the class of Sobolev-Fock vectors we have that $%
a_{s_{0}}\left( T_{s_{0}}^{-}\right) \Phi $ will be norm convergent to $%
a_{r_{0}}\left( T_{r_{0}}^{+}\right) \Phi $ as $\tau _{0}\rightarrow 0^{+}$.
In this limit we therefore identify the values $a_{r_{0}}\left(
T_{r_{0}}^{+}\right) \Phi $ and $a_{s_{0}}\left( T_{s_{0}}^{-}\right) \Phi $
at the start and end of the edge being eliminated. The boundary condition $%
a_{s_{0}}\left( T_{s_{0}}^{-}\right) \Phi =\sum_{\beta \in \left\{ 0\right\}
\cup \mathcal{P}_{\text{in}}}\mathsf{V}_{s_{0}r}a_{r}\left( T_{r}^{+}\right)
\Phi $ may then be rewritten as 
\begin{equation*}
a_{r_{0}}\left( T_{r_{0}}^{+}\right) \Phi =\left( 1-\mathsf{V}%
_{s_{0}r_{0}}\right) ^{-1}\sum_{r\in \left\{ 0\right\} \cup \mathcal{P}_{%
\text{in}}\backslash \left\{ r_{0}\right\} }\mathsf{V}_{s_{0}r}a_{r}\left(
T_{r}^{+}\right) \Phi
\end{equation*}
and substituting into the boundary condition for $s\in \mathcal{P}_{\text{out%
}}\backslash \left\{ s_{0}\right\} $ yields 
\begin{eqnarray*}
a_{s}\left( T_{s}^{-}\right) \Phi &=&\sum_{r\in \left\{ 0\right\} \cup 
\mathcal{P}_{\text{in}}}\mathsf{V}_{sr}a_{r}\left( T_{r}^{+}\right) \Phi \\
&=&\mathsf{V}_{s_{0}0}\Phi +\sum_{r\in \mathcal{P}_{\text{in}}\backslash
\left\{ r_{0}\right\} }\mathsf{V}_{s_{0}r}a_{r}\left( T_{r}^{+}\right) \Phi +%
\mathsf{V}_{s_{0}r_{0}}a_{r_{0}}\left( T_{r}^{+}\right) \Phi \\
&\equiv &\sum_{r\in \left\{ 0\right\} \cup \mathcal{P}_{\text{in}}\backslash
\left\{ r_{0}\right\} }\mathsf{V}_{s_{0}r}^{\text{red}}a_{r}\left(
T_{r}^{+}\right) \Phi
\end{eqnarray*}
The reduced Hamiltonian is then defined by 
\begin{eqnarray*}
-iH^{\text{red}}\Phi &=&\sum_{r\in \left\{ 0\right\} \cup \mathcal{P}_{\text{%
in}}}\mathsf{V}_{0r}a_{r}\left( T_{r}^{+}\right) \Phi -iH_{0}^{\text{red}%
}\Phi \\
&=&\sum_{r\in \left\{ 0\right\} \cup \mathcal{P}_{\text{in}}\backslash
\left\{ r_{0}\right\} }\mathsf{V}_{0r}a_{r}\left( T_{r}^{+}\right) \Phi +%
\mathsf{V}_{0r_{0}}a_{r_{0}}\left( T_{r_{0}}^{+}\right) \Phi -iH_{0}^{\text{%
red}}\Phi \\
&\equiv &\sum_{r\in \left\{ 0\right\} \cup \mathcal{P}_{\text{in}}\backslash
\left\{ r_{0}\right\} }\mathsf{V}_{0r}^{\text{red}}a_{r}\left(
T_{r}^{+}\right) \Phi -iH_{0}^{\text{red}}\Phi ,
\end{eqnarray*}
where $H_{0}^{\text{red}}=\sum_{e\in \mathcal{E}\backslash \left\{
e_{0}\right\} }H_{0}\left( e\right) $.

The above argument rests on the fact that $\left\| a_{r_{0}}\left(
T_{r_{0}}^{+}\right) \Phi -a_{s_{0}}\left( T_{s_{0}}^{-}\right) \Phi
\right\| \rightarrow 0$ as $\tau _{0}\rightarrow 0$ for Sobolev-Fock vectors 
$\Phi $. More explicitly, by rescaling the local time-coordinate on the edge
to have fixed range $\left[ 0,1\right] $, we encounter a sequence of
Hamiltonians $H=H^{\left( \tau _{0}\right) }$ with common domain in $\frak{H}%
_{\mathcal{N}}$. Let $U^{\left( \tau _{0}\right) }$ and $U^{\text{red}}$ be
the one-parameter unitary groups generated by the $H^{\left( \tau
_{0}\right) }$ and $H^{\text{red}}$ respectively. The Hamiltonian $H^{\text{%
red}}$ has domain in $\frak{H}_{\mathcal{N}^{\text{red}}}\subset \frak{H}_{%
\mathcal{N}}$ with the embedding implemented in the obvious way by taking
zero quanta in the eliminated edge. For each $\Phi \in dom\left( H^{\text{red%
}}\right) $ we may construct a sequence $\Phi ^{\left( \tau _{0}\right) }\in
dom\left( H^{\left( \tau _{0}\right) }\right) $ strongly convergent to $\Phi 
$ such that $H^{\left( \tau _{0}\right) }\Phi ^{\left( \tau _{0}\right) }$
converges strongly to $H^{\text{red}}\Phi $. By the Trotter-Kato theorem,
e.g. \cite{Davies} theorem 3.17, we then have that 
\begin{equation*}
\lim_{\tau _{0}\rightarrow 0}\sup_{0\leq t\leq T}\left\| \left(
U_{t}^{\left( \tau _{0}\right) }-U_{t}^{\text{red}}\right) \Phi \right\| =0,
\end{equation*}
for all $T>0$ and $\Phi \in \frak{H}_{\mathcal{N}^{\text{red}}}$.
\end{proof}

\bigskip

The reduced model matrix $\left( \ref{Vred}\right) $\ is closely related to
fractional linear transformations, see for instance \cite{Young}, \cite
{GreenLimebeer}, \cite[Chapter 10]{ZDG96}, and we introduce appropriate
definitions in the present context.

\begin{definition}
Let $\frak{K}_{e}$ be the multiplicity space of an edge $e=\left( s,r\right) 
$ so that $\frak{K}_{e}$ is a subspace of the network multiplicity $\frak{K}%
_{\text{net}}$ and let $X\in B\left( \frak{K}_{e}\right) $. The feedback
reduction of $\mathbf{V}\in \frak{M}\left( \frak{h},\frak{K}\right) $
through the edge $e$, with gain $X$, is the map 
\begin{eqnarray*}
\mathcal{F}_{e} &:&\frak{M}\left( \frak{h},\frak{K}\right) \times B\left( 
\frak{K}_{e}\right) \mapsto \frak{M}\left( \frak{h},\frak{K}\ominus \frak{K}%
_{e}\right) \\
&:&\left( \mathbf{V},X\right) \mapsto \mathcal{F}_{e}\left( \mathbf{V}%
,X\right)
\end{eqnarray*}
where in terms of block decomposition we have 
\begin{equation}
\mathcal{F}_{e}\left( \mathbf{V},X\right) _{\alpha \beta }\triangleq \mathsf{%
V}_{\alpha \beta }+\mathsf{V}_{\alpha r}X\left( 1-\mathsf{V}_{sr}X\right)
^{-1}\mathsf{V}_{s\beta },
\end{equation}
where the indices are $\beta \in \left\{ 0\right\} \cup P_{\text{in}%
}\backslash \left\{ r_{0}\right\} $ and $\alpha \in \left\{ 0\right\} \cup
P_{\text{out}}\backslash \left\{ s_{0}\right\} $. The domain is then the set
of all pairs $\left( \mathbf{V},X\right) $ such that $1-\mathsf{V}_{sr}X$ is
invertible in $B\left( \frak{K}_{e}\right) $. In the special case of unit
gain we write $\mathcal{F}_{e}\left( T\right) \triangleq \mathcal{F}%
_{e}\left( T,1\right) $. For fixed $\mathbf{V}$, the map $\mathcal{F}%
_{e}\left( \mathbf{V},\cdot \right) $ is a \textit{non-commutative
fractional linear, or M\"{o}bius, transformation}.
\end{definition}

\begin{lemma}[Siegel]
If $S=\left( 
\begin{array}{cc}
A & B \\ 
C & D
\end{array}
\right) $ be a unitary operator on the direct sum $\frak{H}_{1}\oplus \frak{H%
}_{2}$ of two Hilbert spaces with $\left\| A\right\| <1$. Let $\Phi
_{T}\left( X\right) \triangleq D+CX\left( 1-AX\right) ^{-1}B$ with $X\in
Dom(\Phi _{S})$ whenever $\left\| X\right\| \leq 1$. For $X,Y\in Dom(\Phi
_{S})$ we have the \textit{Siegel identities}. 
\begin{eqnarray*}
\Phi _{S}\left( X\right) ^{\dag }\,\Phi _{S}\left( Y\right) -1 &=&B^{\dag
}\left( 1-X^{\dag }A^{\dag }\right) ^{-1}\left( X^{\dag }Y-1\right) \left(
1-AY\right) B, \\
\Phi _{S}\left( X\right) \,\Phi _{S}\left( Y\right) ^{\dag }-1 &=&C^{\dag
}\left( 1-XA\right) ^{-1}\left( XY^{\dag }-1\right) \left( 1-A^{\dag
}Y\right) C^{\dag }.
\end{eqnarray*}
\end{lemma}

A proof can be found in \cite{Young}.

\begin{corollary}
If $S=\left( 
\begin{array}{cc}
A & B \\ 
C & D
\end{array}
\right) $ is unitary and $1-A$ is invertible, then $D+C\left( 1-A\right)
^{-1}B$ is unitary.
\end{corollary}

For consistency, we would hope that the reduced model matrix belongs to the
class of model matrices with multiplicity space one dimension lower. This we
now show to be the case.

\begin{lemma}
Let $\mathbf{V}$ be the model matrix determined by the operators $\left( 
\mathsf{S},\mathsf{L},\mathsf{H}\right) $. Then the reduced model matrix $%
\mathbf{V}^{\text{red}}$ obtained by eliminating the edge $\left(
r_{0},s_{0}\right) $ is determined by the operators $\left( \mathsf{S}^{%
\text{red}},\mathsf{L}^{\text{red}},\mathsf{H}^{\text{red}}\right) $\ where 
\begin{eqnarray*}
\mathsf{S}_{sr}^{\text{red}} &=&\mathsf{S}_{sr}+\mathsf{S}_{sr_{0}}\left( 1-%
\mathsf{S}_{s_{0}r_{0}}\right) ^{-1}\mathsf{S}_{s_{0}r}, \\
\mathsf{L}_{s}^{\text{red}} &=&\mathsf{L}_{s}+\mathsf{S}_{sr_{0}}\left( 1-%
\mathsf{S}_{s_{0}r_{0}}\right) ^{-1}\mathsf{L}_{s_{0}}, \\
\mathsf{H}^{\text{red}} &=&\mathsf{H}+\sum_{s\in \mathcal{P}_{\text{out}}}%
\func{Im}\mathsf{L}_{s}^{\dag }\mathsf{S}_{sr_{0}}\left( 1-\mathsf{S}%
_{s_{0}r_{0}}\right) ^{-1}\mathsf{L}_{s_{0}},
\end{eqnarray*}
for $r\in \mathcal{P}_{\text{in}}\backslash \left\{ r_{0}\right\} $ and $%
s\in \mathcal{P}_{\text{out}}\backslash \left\{ s_{0}\right\} $.
\end{lemma}

\begin{proof}
The identifications $\mathsf{V}_{sr}^{\text{red}}=\mathsf{S}_{sr}^{\text{red}%
}=\mathsf{S}_{sr}+\mathsf{S}_{sr_{0}}\left( 1-\mathsf{S}_{s_{0}r_{0}}\right)
^{-1}\mathsf{S}_{s_{0}r}$ and $\mathsf{V}_{s0}^{\text{red}}=\mathsf{L}_{s}^{%
\text{red}}=\mathsf{L}_{s}+\mathsf{S}_{sr_{0}}\left( 1-\mathsf{S}%
_{s_{0}r_{0}}\right) ^{-1}\mathsf{L}_{s_{0}}$ are immediate. Unitarity of $%
\mathsf{S}^{\text{red}}$\ follows from the above corollary to the Siegel
identities. We next check that $\mathsf{V}_{0r}^{\text{red}}\equiv
-\sum_{s\in \mathcal{P}_{\text{out}}\backslash \left\{ s_{0}\right\} }%
\mathsf{L}_{s}^{\text{red}\dag }\mathsf{S}_{sr}^{\text{red}}$. Here 
\begin{equation*}
\sum_{s\in \mathcal{P}_{\text{out}}\backslash \left\{ s_{0}\right\} }\mathsf{%
L}_{s}^{\text{red}\dag }\mathsf{S}_{sr}^{\text{red}}=\sum_{s\in \mathcal{P}_{%
\text{out}}\backslash \left\{ s_{0}\right\} }\mathsf{L}_{s}^{\dag }\mathsf{S}%
_{sr}^{\text{red}}+\sum_{s\in \mathcal{P}_{\text{out}}\backslash \left\{
s_{0}\right\} }\mathsf{L}_{s_{0}}^{\dag }\left( 1-\mathsf{S}%
_{s_{0}r_{0}}^{\dag }\right) ^{-1}\mathsf{S}_{sr_{0}}^{\dag }\mathsf{S}%
_{sr}^{\text{red}}
\end{equation*}
and we use the simplification 
\begin{eqnarray*}
\sum_{s\in \mathcal{P}_{\text{out}}\backslash \left\{ s_{0}\right\} }\mathsf{%
S}_{sr_{0}}^{\dag }\mathsf{S}_{sr}^{\text{red}} &=&\sum_{s\in \mathcal{P}_{%
\text{out}}\backslash \left\{ s_{0}\right\} }\left( \mathsf{S}%
_{sr_{0}}^{\dag }\mathsf{S}_{sr}+\mathsf{S}_{sr_{0}}^{\dag }\mathsf{S}%
_{sr_{0}}\left( 1-\mathsf{S}_{s_{0}r_{0}}\right) ^{-1}\mathsf{S}%
_{s_{0}r}\right) \\
&=&\left( -\mathsf{S}_{s_{0}r_{0}}^{\dag }\mathsf{S}_{s_{0}r}+(1-\mathsf{S}%
_{s_{0}r_{0}}^{\dag }\mathsf{S}_{s_{0}r_{0}})\left( 1-\mathsf{S}%
_{s_{0}r_{0}}\right) ^{-1}\mathsf{S}_{s_{0}r}\right) \\
&=&\left( -\mathsf{S}_{s_{0}r_{0}}^{\dag }(1-\mathsf{S}_{s_{0}r_{0}})+(1-%
\mathsf{S}_{s_{0}r_{0}}^{\dag }\mathsf{S}_{s_{0}r_{0}})\right) \left( 1-%
\mathsf{S}_{s_{0}r_{0}}\right) ^{-1}\mathsf{S}_{s_{0}r} \\
&=&(1-\mathsf{S}_{s_{0}r_{0}}^{\dag })\left( 1-\mathsf{S}_{s_{0}r_{0}}%
\right) ^{-1}\mathsf{S}_{s_{0}r}
\end{eqnarray*}
so that 
\begin{equation*}
\sum_{s\in \mathcal{P}_{\text{out}}\backslash \left\{ s_{0}\right\} }\mathsf{%
L}_{s}^{\text{red}\dag }\mathsf{S}_{sr}^{\text{red}}=\sum_{s\in \mathcal{P}_{%
\text{out}}\backslash \left\{ s_{0}\right\} }\mathsf{L}_{s}^{\dag }\mathsf{S}%
_{sr}^{\text{red}}+\mathsf{L}_{s_{0}}^{\dag }\left( 1-\mathsf{S}%
_{s_{0}r_{0}}\right) ^{-1}\mathsf{S}_{s_{0}r}.
\end{equation*}
On the other hand 
\begin{eqnarray*}
\mathsf{V}_{0r}^{\text{red}} &=&\mathsf{V}_{0r}+\mathsf{V}_{0r_{0}}\left( 1-%
\mathsf{V}_{s_{0}r_{0}}\right) ^{-1}\mathsf{V}_{s_{0}r} \\
&=&-\sum_{s\in \mathcal{P}_{\text{out}}}\mathsf{L}_{s}^{\dag }\mathsf{S}%
_{sr}-\sum_{s\in \mathcal{P}_{\text{out}}}\mathsf{L}_{s}^{\dag }\mathsf{S}%
_{sr_{0}}\left( 1-\mathsf{S}_{s_{0}r_{0}}\right) ^{-1}\mathsf{S}_{s_{0}r} \\
&=&-\sum_{s\in \mathcal{P}_{\text{out}}\backslash \left\{ s_{0}\right\} }%
\mathsf{L}_{s}^{\dag }\mathsf{S}_{sr}^{\text{red}}-\mathsf{L}_{s_{0}}^{\dag }%
\mathsf{S}_{s_{0}r}-\mathsf{L}_{s_{0}}^{\dag }\mathsf{S}_{s_{0}r_{0}}\left(
1-\mathsf{S}_{s_{0}r_{0}}\right) ^{-1}\mathsf{S}_{s_{0}r} \\
&=&-\sum_{s\in \mathcal{P}_{\text{out}}\backslash \left\{ s_{0}\right\} }%
\mathsf{L}_{s}^{\dag }\mathsf{S}_{sr}^{\text{red}}-\mathsf{L}_{s_{0}}^{\dag
}\left( 1-\mathsf{S}_{s_{0}r_{0}}\right) ^{-1}\mathsf{S}_{s_{0}r}\equiv
-\sum_{s\in \mathcal{P}_{\text{out}}\backslash \left\{ s_{0}\right\} }%
\mathsf{L}_{s}^{\text{red}\dag }\mathsf{S}_{sr}^{\text{red}}.
\end{eqnarray*}
Finally we must check that $\mathsf{V}_{00}^{\text{red}}=-\frac{1}{2}%
\sum_{s\in \mathcal{P}_{\text{out}}\backslash \left\{ s_{0}\right\} }\mathsf{%
L}_{s}^{\text{red}\dag }\mathsf{L}_{s}^{\text{red}}-i\mathsf{H}^{\text{red}}$%
. Let us use this as the definition of $\mathsf{H}^{\text{red}}$ then we
have 
\begin{eqnarray*}
-i\mathsf{H}^{\text{red}} &=&\mathsf{V}_{00}+\mathsf{V}_{0r_{0}}\left( 1-%
\mathsf{V}_{s_{0}r_{0}}\right) ^{-1}\mathsf{V}_{s_{0}0}+\frac{1}{2}%
\sum_{s\in \mathcal{P}_{\text{out}}\backslash \left\{ s_{0}\right\} }\mathsf{%
L}_{s}^{\text{red}\dag }\mathsf{L}_{s}^{\text{red}} \\
&=&(-i\mathsf{H}-\frac{1}{2}\sum_{s\in \mathcal{P}_{\text{out}}}\mathsf{L}%
_{s}^{\dag }\mathsf{L}_{s})-\sum_{s\in \mathcal{P}_{\text{out}}}\mathsf{L}%
_{s}^{\dag }\mathsf{S}_{sr_{0}}\left( 1-\mathsf{S}_{s_{0}r_{0}}\right) ^{-1}%
\mathsf{L}_{s_{0}} \\
&&+\frac{1}{2}\sum_{s\in \mathcal{P}_{\text{out}}\backslash \left\{
s_{0}\right\} }\mathsf{L}_{s}^{\text{red}\dag }\mathsf{L}_{s}^{\text{red}}
\end{eqnarray*}
\begin{eqnarray*}
&=&-i\mathsf{H}-\frac{1}{2}\mathsf{L}_{s_{0}}^{\dag }\mathsf{L}_{s_{0}}-%
\frac{1}{2}\sum_{s\in \mathcal{P}_{\text{out}}\backslash \left\{
s_{0}\right\} }\mathsf{L}_{s}^{\dag }\mathsf{S}_{sr_{0}}\left( 1-\mathsf{S}%
_{s_{0}r_{0}}\right) ^{-1}\mathsf{L}_{s_{0}} \\
&&+\frac{1}{2}\sum_{s\in \mathcal{P}_{\text{out}}\backslash \left\{
s_{0}\right\} }\mathsf{L}_{s_{0}}^{\dag }\left( 1-\mathsf{S}%
_{s_{0}r_{0}}^{\dag }\right) ^{-1}\mathsf{S}_{sr_{0}}\mathsf{L}_{s}-\mathsf{L%
}_{s_{0}}^{\dag }\mathsf{S}_{s_{0}r_{0}}\left( 1-\mathsf{S}%
_{s_{0}r_{0}}\right) ^{-1}\mathsf{L}_{s_{0}} \\
&&+\frac{1}{2}\sum_{s\in \mathcal{P}_{\text{out}}\backslash \left\{
s_{0}\right\} }\mathsf{L}_{s_{0}}^{\dag }\left( 1-\mathsf{S}%
_{s_{0}r_{0}}^{\dag }\right) ^{-1}\mathsf{S}_{sr_{0}}^{\dag }\mathsf{S}%
_{sr_{0}}\left( 1-\mathsf{S}_{s_{0}r_{0}}\right) ^{-1}\mathsf{L}_{s_{0}}.
\end{eqnarray*}
We collected together several terms to get $-\frac{1}{2}\mathsf{L}%
_{s_{0}}^{\dag }\mathsf{XL}_{s_{0}}$ where 
\begin{eqnarray*}
\mathsf{X} &=&1+2\mathsf{S}_{s_{0}r_{0}}\left( 1-\mathsf{S}%
_{s_{0}r_{0}}\right) ^{-1}-\sum_{s\in \mathcal{P}_{\text{out}}\backslash
\left\{ s_{0}\right\} }\left( 1-\mathsf{S}_{s_{0}r_{0}}^{\dag }\right) ^{-1}%
\mathsf{S}_{sr_{0}}^{\dag }\mathsf{S}_{sr_{0}}\left( 1-\mathsf{S}%
_{s_{0}r_{0}}\right) ^{-1} \\
&=&1+2\mathsf{S}_{s_{0}r_{0}}\left( 1-\mathsf{S}_{s_{0}r_{0}}\right)
^{-1}-\left( 1-\mathsf{S}_{s_{0}r_{0}}^{\dag }\right) ^{-1}\left( 1-\mathsf{S%
}_{s_{0}r_{0}}^{\dag }\mathsf{S}_{s_{0}r_{0}}\right) \left( 1-\mathsf{S}%
_{s_{0}r_{0}}\right) ^{-1} \\
&=&\left( 1-\mathsf{S}_{s_{0}r_{0}}^{\dag }\right) ^{-1}\left( \mathsf{S}%
_{s_{0}r_{0}}-\mathsf{S}_{s_{0}r_{0}}^{\dag }\right) \left( 1-\mathsf{S}%
_{s_{0}r_{0}}\right) ^{-1}.
\end{eqnarray*}
It follows that 
\begin{equation*}
\mathsf{H}^{\text{red}}=\mathsf{H}+\func{Im}\sum_{s\in \mathcal{P}_{\text{out%
}}\backslash \left\{ s_{0}\right\} }\mathsf{L}_{s}^{\dag }\mathsf{S}%
_{sr_{0}}\left( 1-\mathsf{S}_{s_{0}r_{0}}\right) ^{-1}\mathsf{L}_{s_{0}}+%
\func{Im}\mathsf{L}_{s_{0}}^{\dag }\left( 1-\mathsf{S}_{s_{0}r_{0}}\right)
^{-1}\mathsf{L}_{s_{0}}
\end{equation*}
Note that $\func{Im}\mathsf{L}_{s_{0}}^{\dag }\left( 1-\mathsf{S}%
_{s_{0}r_{0}}\right) ^{-1}\mathsf{L}_{s_{0}}=\func{Im}\mathsf{L}%
_{s_{0}}^{\dag }\mathsf{S}_{s_{0}r_{0}}\left( 1-\mathsf{S}%
_{s_{0}r_{0}}\right) ^{-1}\mathsf{L}_{s_{0}}$ we obtain $\mathsf{H}^{\text{%
red}}$ as the self-adjoint operator given in the statement of the lemma.
\end{proof}

\bigskip

\begin{lemma}
Let $e_{1}=\left( s_{1},r_{1}\right) $ and $e_{2}=\left( s_{2},r_{2}\right) $
be a pair of edges in a network then 
\begin{equation}
\mathcal{F}_{e_{1}}\circ \mathcal{F}_{e_{2}}=\mathcal{F}_{e_{2}}\circ 
\mathcal{F}_{e_{1}}=\mathcal{F}_{e_{1}\oplus e_{2}}.
\end{equation}
\end{lemma}

\begin{proof}
Without loss of generality suppose that $e_{1}=\left( 1,1\right) $ and $%
e_{2}=\left( 2,2\right) $, then for $\alpha ,\beta \notin \left\{
1,2\right\} $%
\begin{equation*}
(\mathcal{F}_{e_{2}}\circ \mathcal{F}_{e_{1}}\mathbf{V)}_{\alpha \beta }=%
\mathsf{\hat{V}}_{\alpha \beta }+\mathsf{\hat{V}}_{\alpha 2}\left( 1-\mathsf{%
\hat{V}}_{22}\right) ^{-1}\mathsf{\hat{V}}_{2\beta }
\end{equation*}
where $\mathsf{\hat{V}}_{\alpha \beta }=\mathsf{V}_{\alpha \beta }+\mathsf{V}%
_{\alpha 1}\left( 1-\mathsf{V}_{11}\right) ^{-1}\mathsf{V}_{1\beta }$, and
so we have 
\begin{equation*}
(\mathcal{F}_{e_{2}}\circ \mathcal{F}_{e_{1}}\mathbf{V)}_{\alpha \beta }=%
\mathsf{\hat{V}}_{\alpha \beta }+ 
\begin{array}{c}
(\mathsf{V}_{\alpha 1},\mathsf{V}_{\alpha 2}) \\ 
\text{ }
\end{array}
Z\left( 
\begin{array}{c}
\mathsf{V}_{\alpha 1} \\ 
\mathsf{V}_{\alpha 2}
\end{array}
\right)
\end{equation*}
with 
\begin{eqnarray*}
Z &=&\left( 
\begin{array}{cc}
\frac{1}{1-\mathsf{V}_{11}}+\frac{1}{1-\mathsf{V}_{11}}\mathsf{V}_{12}\frac{1%
}{1-\mathsf{\hat{V}}_{22}}\mathsf{V}_{21}\frac{1}{1-\mathsf{V}_{11}} & \frac{%
1}{1-\mathsf{V}_{11}}\mathsf{V}_{12}\frac{1}{1-\mathsf{\hat{V}}_{22}} \\ 
\frac{1}{1-\mathsf{\hat{V}}_{22}}\mathsf{V}_{21}\frac{1}{1-\mathsf{V}_{11}}
& \frac{1}{1-\mathsf{\hat{V}}_{22}}
\end{array}
\right) \\
&\equiv &\left( 
\begin{array}{cc}
1-\mathsf{V}_{11} & -\mathsf{V}_{12} \\ 
-\mathsf{V}_{21} & 1-\mathsf{V}_{22}
\end{array}
\right) ^{-1}.
\end{eqnarray*}
The expression is clearly symmetric under interchange of$\ 1$ and $2$, and
corresponds to the double edge elimination.
\end{proof}

This implies that the order in which we apply zero time delay limits to
eliminate multiple internal channels does not in fact matter, and can be
combined simultaneously. In this manner, every quantum network may be
reduced to a single Markovian component in a unique well-defined
algebraically manner by eliminating the time delays in all internal channels
by means of the map 
\begin{equation*}
\mathcal{F}=\circ _{e\in \mathcal{E}_{\text{int}}}\mathcal{F}_{e}:\frak{M}%
\left( \frak{h},\frak{K}_{\text{net}}\right) \mapsto \frak{M}\left( \frak{h},%
\frak{K}_{\text{ext}}\right) .
\end{equation*}

If we decide to eliminate all internal channels by making the connections,
inserting a gain matrix $X$, and taking the zero time-delay limit, then the
resulting network will have model matrix $\mathcal{F}\left( \mathbf{V}%
,X\right) $ and of course has dimensions determined by the remaining
(external) channels. The functorial property is that we are able to
eliminate blocks of channels in one step, giving the same answer as if we
performed the eliminations one-by-one.

\section{Physical Applications}

We shall be interested in the situation where we eliminate \textit{all} the 
\textit{internal channels} (total multiplicity space $\frak{K}_{\mathtt{i}}$%
) leaving only the external channels (total multiplicity space $\frak{K}_{%
\mathtt{e}}$). With respect to the decomposition $\frak{K}=\frak{K}_{\mathtt{%
int}}\oplus \frak{K}_{\mathtt{ext}}$, we write the operators $\left( \mathsf{%
S},\mathsf{L},\mathsf{H}\right) $ of the network as 
\begin{equation*}
\mathsf{S}=\left( 
\begin{array}{cc}
\mathsf{S}_{\mathtt{ii}} & \mathsf{S}_{\mathtt{ie}} \\ 
\mathsf{S}_{\mathtt{ei}} & \mathsf{S}_{\mathtt{ee}}
\end{array}
\right) ,\quad \mathsf{L}=\left( 
\begin{array}{c}
\mathsf{L}_{\mathtt{i}} \\ 
\mathsf{L}_{\mathtt{e}}
\end{array}
\right) .
\end{equation*}
The feedback reduced model matrix may be conveniently expressed as 
\begin{equation*}
\mathcal{F}\left( \mathbf{V,\eta }^{-1}\right) _{\alpha \beta }\triangleq 
\mathsf{V}_{\alpha \beta }+\mathsf{V}_{\alpha \mathtt{i}}\left( \eta -%
\mathsf{V}_{\mathtt{ii}}\right) ^{-1}\mathsf{V}_{\mathtt{i}\beta }
\end{equation*}
for $\alpha ,\beta \in \left\{ 0,\mathtt{e}\right\} $, where $\eta $ is the
(unitary) \textit{adjacency matrix} 
\begin{equation*}
\eta _{sr}=\left\{ 
\begin{array}{cc}
1, & \text{if }\left( s,r\right) \text{ is an internal channel,} \\ 
0, & \text{otherwise.}
\end{array}
\right.
\end{equation*}
Here the ``gain'' $\eta $ is the set of instructions as to which internal
output port gets connected up to which internal input port. We, of course,
have $\eta =1$ if we match up the labels of the input and output ports
according to the connections, however, it is computationally easier to work
with a general labelling and just specify the adjacency matrix. The reduced
model matrix $\mathbf{V}^{\text{red}}$ obtained by eliminating all the
internal channels is determined by the operators $\left( \mathsf{S}^{\text{%
red}},\mathsf{L}^{\text{red}},\mathsf{H}^{\text{red}}\right) $\ given by 
\begin{eqnarray*}
\mathsf{S}^{\text{red}} &=&\mathsf{S}_{\mathtt{ee}}+\mathsf{S}_{\mathtt{ei}%
}\left( \eta -\mathsf{S}_{\mathtt{ii}}\right) ^{-1}\mathsf{S}_{\mathtt{ie}},
\\
\mathsf{L}^{\text{red}} &=&\mathsf{L}_{\mathtt{e}}+\mathsf{S}_{\mathtt{ei}%
}\left( \eta -\mathsf{S}_{\mathtt{ii}}\right) ^{-1}\mathsf{L}_{\mathtt{i}},
\\
\mathsf{H}^{\text{red}} &=&\mathsf{H}+\sum_{i=\mathtt{i},\mathtt{e}}\func{Im}%
\mathsf{L}_{j}^{\dag }\mathsf{S}_{j\mathtt{i}}\left( \eta -\mathsf{S}_{%
\mathtt{ii}}\right) ^{-1}\mathsf{L}_{\mathtt{i}}.
\end{eqnarray*}

\subsection{Quantum Systems in Feedforward}

Our first application is to derive the formula for systems in series. This
is the simplest nontrivial example of a quantum network.

\begin{center}
%
%
%
%
%
%
%
%
%
%
%
\setlength{\unitlength}{.1cm}
\begin{picture}(100,22)
\label{pic3}

\thicklines

\put(30,5){\line(0,1){10}}
\put(30,5){\line(1,0){20}}
\put(50,5){\line(0,1){10}}
\put(30,15){\line(1,0){20}}

\put(60,5){\line(0,1){10}}
\put(60,5){\line(1,0){20}}
\put(80,5){\line(0,1){10}}
\put(60,15){\line(1,0){20}}

\thinlines
\put(32,10){\vector(-1,0){15}}
\put(62,10){\vector(-1,0){14}}

\put(92,10){\vector(-1,0){14}}

\put(33,10){\circle{2}}
\put(63,10){\circle{2}}

\put(47,10){\circle{2}}
\put(77,10){\circle{2}}

\put(35,11){$s_2$}
\put(65,11){$s_1$}
\put(42,11){$r_2$}
\put(72,11){$r_1$}

\end{picture}%
%

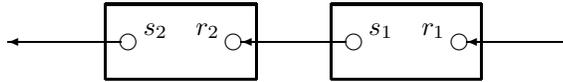
figure 3: Systems in series
\end{center}

We begin by taking $\left( \mathsf{S}_{1},\mathsf{L}_{1},\mathsf{H}%
_{1}\right) $ and $\left( \mathsf{S}_{2},\mathsf{L}_{2},\mathsf{H}%
_{2}\right) $ to be the operators of the first and second system when
considered as separate systems driven by independent noises. The model
matrix for the network is then 
\begin{equation*}
\mathbf{V}=\left( 
\begin{array}{ccc}
-\sum_{j=1,2}(\frac{1}{2}\mathsf{L}_{j}^{\dag }\mathsf{L}_{j}+i\mathsf{H}%
_{j}) & -\mathsf{L}_{1}^{\dag }\mathsf{S}_{1} & -\mathsf{L}_{2}^{\dag }%
\mathsf{S}_{2} \\ 
\mathsf{L}_{1} & \mathsf{S}_{1} & 0 \\ 
\mathsf{L}_{2} & 0 & \mathsf{S}_{2}
\end{array}
\right)
\end{equation*}
with respect to the labelling $s=\{0,s_{1},s_{2}\}$ for the outputs (rows)
and $r=\left\{ 0,r_{1},r_{2}\right\} $ for the inputs (columns). We wish to
reform a feedback reduction wherein we connect the systems via edge $%
e=\left( s_{1},r_{2}\right) $ and take the zero time-delay limit along the
edge. The resulting model should then be Markovian and its model matrix is
given by 
\begin{equation*}
\mathbf{V}_{\text{series}}=\mathcal{F}_{e}\mathbf{V}
\end{equation*}
\begin{eqnarray*}
&=&\left( 
\begin{array}{cc}
-\sum_{j=1,2}(\frac{1}{2}\mathsf{L}_{j}^{\dag }\mathsf{L}_{j}+i\mathsf{H}%
_{j}) & -\mathsf{L}_{1}^{\dag }\mathsf{S}_{1} \\ 
\mathsf{L}_{2} & 0
\end{array}
\right) +\left( 
\begin{array}{c}
-\mathsf{L}_{2}^{\dag }\mathsf{S}_{2} \\ 
\mathsf{S}_{2}
\end{array}
\right) \left( 1-0\right) ^{-1} 
\begin{array}{c}
\left( \mathsf{L}_{1},\mathsf{S}_{1}\right) \\ 
\,
\end{array}
\\
&=&\left( 
\begin{array}{cc}
-\sum_{j=1,2}(\frac{1}{2}\mathsf{L}_{j}^{\dag }\mathsf{L}_{j}+i\mathsf{H}%
_{j})-\mathsf{L}_{2}^{\dag }\mathsf{S}_{2}\mathsf{L}_{1} & -\mathsf{L}%
_{1}^{\dag }\mathsf{S}_{1}-\mathsf{L}_{2}^{\dag }\mathsf{S}_{2}\mathsf{S}_{1}
\\ 
\mathsf{L}_{2}+\mathsf{S}_{2}\mathsf{L}_{1} & \mathsf{S}_{2}\mathsf{S}_{1}
\end{array}
\right) ,
\end{eqnarray*}
that is, the operators determining the reduced system are 
\begin{eqnarray*}
\mathsf{S}_{\text{series}} &=&\mathsf{S}_{2}\mathsf{S}_{1}, \\
\mathsf{L}_{\text{series}} &=&\mathsf{L}_{2}+\mathsf{S}_{2}\mathsf{L}_{1}, \\
\mathsf{H}_{\text{series}} &=&\mathsf{H}_{1}+\mathsf{H}_{2}+\func{Im}\left\{ 
\mathsf{L}_{2}^{\dag }\mathsf{S}_{2}\mathsf{L}_{1}\right\} .
\end{eqnarray*}
The Evans-Hudson maps associated with the feedforward system can be related
to those of the individual systems via the identity 
\begin{equation}
\mathcal{L}_{\alpha \beta }\left( \cdot \right) =\mathcal{L}_{\alpha \beta
}^{\left( 1\right) }\left( \cdot \right) +(\mathsf{M}_{\mu \alpha }^{\left(
1\right) })^{\dag }\mathcal{L}_{\mu \nu }^{\left( 2\right) }\left( \cdot
\right) \mathsf{M}_{\nu \beta }^{\left( 1\right) }.  \label{EH in series}
\end{equation}

The case most familiar to the quantum optics community is two cavity systems
in cascade. Here \textsf{$S$}$_{1}=$\textsf{$S$}$_{2}=1$ and \textsf{$L$}$%
_{i}=\sqrt{\gamma _{i}}a_{i}$ $\left( i=1,2\right) $ so we obtain \textsf{$L$%
}$=\sqrt{\gamma _{1}}a_{1}+\sqrt{\gamma _{2}}a_{2},$ $\mathsf{H}=\mathsf{H}%
_{1}+\mathsf{H}_{2}+\frac{1}{2i}\sqrt{\gamma _{1}\gamma _{2}}\left(
a_{2}^{\dag }a_{1}-a_{1}^{\dag }a_{2}\right) $. This agrees with the
calculations of Gardiner \cite{Gardiner} for cascaded oscillators.
Gardiner's derivation \cite{Gardiner} of the cascade rule is different from
ours, but would extend to the cover the gauge case due to $\left( \ref{EH in
series}\right) $.

\subsubsection{The Series Product}

The rule for determining the form of the model for\ systems in series has
been previously given for the It\={o} generator matrices where it was called
the \textit{series product}. It is related to the general question of how to
``add'' stochastic derivations in order to obtain a stochastic derivation 
\cite{AH}. We recall its definition and establish its basic properties.

\begin{definition}
Let $\mathbf{G}_{1}$ and $\mathbf{G}_{2}$ be the It\={o} generator matrices
with the same multiplicity spaces, then the series product is defined to be 
\begin{equation*}
\mathbf{G}_{2}\vartriangleleft \mathbf{G}_{1}\triangleq \mathbf{G}_{1}+%
\mathbf{G}_{2}+\mathbf{G}_{2}\mathbf{\Pi G}_{1}.
\end{equation*}
\end{definition}

\begin{lemma}
The series product of two It\={o} matrices $\mathbf{G}=\mathbf{G}%
_{2}\vartriangleleft \mathbf{G}_{1}$ is again an It\={o} matrix and if the $%
\mathbf{G}_{i}$ have parameters $\left( \mathsf{S}_{i},\mathsf{L}_{i},%
\mathsf{H}\right) $ then $\mathbf{G}$ has parameters $\left( \mathsf{S}_{%
\text{series}},\mathsf{L}_{\text{series}},\mathsf{H}_{\text{series}}\right) $%
. If $\mathbf{M}_{i}$ is the Galilean matrix associated with $\mathbf{G}_{i}$
then the Galilean matrix associated with $\mathbf{G}=\mathbf{G}%
_{2}\vartriangleleft \mathbf{G}_{1}$ is $\mathbf{M}=\mathbf{M}_{2}\mathbf{M}%
_{1}$. The series product is not symmetric, but is associative.
\end{lemma}

\begin{proof}
One readily checks that the series product of two It\={o} matrices satisfies
the conditions to be an It\={o} generating matrix. The specific form of the
parameters is found by inspection. The associated Galilean transformation is 
\begin{eqnarray*}
\mathbf{M} &=&\mathbf{1}+\mathbf{\Pi }\left( \mathbf{G}_{1}+\mathbf{G}_{2}+%
\mathbf{G}_{2}\mathbf{\Pi G}_{1}\right) \\
&=&\mathbf{1}+\mathbf{\Pi G}_{1}+\mathbf{\Pi G}_{2}+\mathbf{\Pi G}_{2}%
\mathbf{\Pi G}_{1} \\
&=&\left( \mathbf{1}+\mathbf{\Pi G}_{2}\right) \left( \mathbf{1}+\mathbf{\Pi
G}_{1}\right) \\
&=&\mathbf{M}_{2}\mathbf{M}_{1}.
\end{eqnarray*}
To prove associativity, let us construct the $\left( 1+n+1\right) $-square
matrix 
\begin{equation}
\mathbb{V}=\left( 
\begin{array}{ccc}
1 & \mathsf{-L}^{\dag }\mathsf{S} & \mathsf{-}\frac{1}{2}\mathsf{L}^{\dag }%
\mathsf{L}-i\mathsf{H} \\ 
0 & \mathsf{S} & \mathsf{L} \\ 
0 & 0 & 1
\end{array}
\right)
\end{equation}
from the model parameters $\left( \mathsf{S},\mathsf{L},\mathsf{H}\right) $.
Then the series product corresponds to the ordinary matrix product $\mathbb{V%
}_{\text{series}}=\mathbb{V}_{2}\mathbb{V}_{2}$ which is clearly associative.
\end{proof}

\bigskip

Associativity means that we can extend the result immediately to several
systems cascaded in series. The easiest way to calculate the model matrix
for several components in series is then by the ordinary matrix product of
the augmented matrices 
\begin{equation*}
\mathbb{V}_{\text{series}}=\mathbb{V}_{n}\cdots \mathbb{V}_{2}\mathbb{V}_{1}.
\end{equation*}
In the lemma, the matrix $\mathbb{V}$ is of the type introduced by Belavkin 
\cite{Belavkin98} to efficiently capture the It\={o} correction as an
ordinary product. Let $\zeta =\left( 
\begin{array}{ccc}
0 & 0 & 1 \\ 
0 & 1 & 0 \\ 
1 & 0 & 0
\end{array}
\right) $ then we obtain an involution $\mathbb{X}^{\star }=\zeta \mathbb{X}%
^{\dag }\zeta $ on the space of $\left( 1+n+1\right) $-dimensional matrices.
Then we are considering precisely the class of $\star $-unitary matrices of
the form $\mathbb{V}=\left( 
\begin{array}{ccc}
1 & B & A \\ 
0 & D & C \\ 
0 & 0 & 1
\end{array}
\right) $, that is $\mathbb{V}^{\star }\mathbb{V}=\mathbb{VV}^{\star }=1$.
In particular, the product of two $\star $-unitaries is again a $\star $%
-unitary.

\begin{remark}
Finally let us make the important remark that nowhere did we assume that the
entries of $\mathbf{G}_{1}$ and $\mathbf{G}_{2}$ had to commute. This means
that the series product can describe not only forward from one component
system to an independent system, but also feedback into itself as well. This
is captured in the picture below.
\end{remark}

\subsection{Beam Splitters}

A simple beam splitter is a device performing physical superposition of two
input fields. It is described by a fixed unitary operator $T=\left( 
\begin{array}{cc}
\alpha  & \beta  \\ 
\mu  & \nu 
\end{array}
\right) \in U\left( 2\right) $: 
\begin{equation*}
\left( 
\begin{array}{c}
\tilde{A}_{1} \\ 
\tilde{A}_{2}
\end{array}
\right) =\left( 
\begin{array}{cc}
\alpha  & \beta  \\ 
\mu  & \nu 
\end{array}
\right) \left( 
\begin{array}{c}
A_{1} \\ 
A_{2}
\end{array}
\right) .
\end{equation*}
This is a canonical transformation and the output fields satisfy the same
canonical commutation relations as the inputs. The action of the beam
splitter is depicted in the figure below. On the left we have a traditional
view of the two inputs $\left( A_{1},A_{2}\right) $ being split into two
output fields $(\tilde{A}_{1},\tilde{A}_{2})$. On the right we have our view
of the beam splitter as being a component with two input ports and two
output ports: we have sketched some internal detail to emphasize how the
scattering (superimposing) of inputs however we shall usually just draw this
as a ``black box'' component in the following.

\begin{center}
%
%
%
%
%
%
%
%
%
\setlength{\unitlength}{.075cm}
\begin{picture}(100,40)
\label{pic4}
\thinlines

\put(64,10){\vector(-1,0){9}}
\put(64,25){\vector(-1,0){9}}
\put(66,10){\line(1,0){13}}
\put(66,25){\line(1,0){13}}
\put(90,10){\vector(-1,0){9}}
\put(90,25){\vector(-1,0){9}}
\put(66,11){\line(1,1){13}}
\put(66,24){\line(1,-1){13}}

\put(65,10){\circle{2}}
\put(65,25){\circle{2}}
\put(80,10){\circle{2}}
\put(80,25){\circle{2}}

\thinlines
\put(17,20){\vector(1,0){12}}
\put(15,5){\vector(0,1){12}}
\put(0,20){\vector(1,0){12}}
\put(15,23){\vector(0,1){12}}
\thicklines
\put(7,12){\line(1,1){16}}

\put(88,12){$A_2$}
\put(88,27){$A_1$}
\put(48,12){${\tilde A}_2$}
\put(48,27){${\tilde A}_1$}

\put(18,8){$A_2$}
\put(0,22){$A_1$}
\put(17,35){${\tilde A}_2$}
\put(30,22){${\tilde A}_1$}

\put(62,7){\dashbox(22,20)}

\end{picture}
%

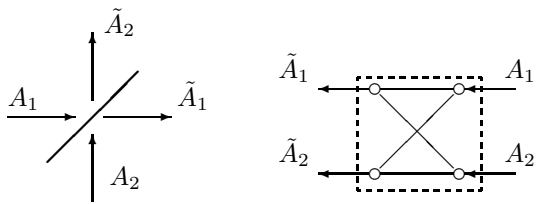
Figure 4: Beam-splitter component.
\end{center}

Our aim is to describe the effective Markov model for the feedback device
sketched below where the feedback is achieved by means of a beam splitter.
Here we have a component system, called the plant, in-loop and we assume
that it is described by the parameters $\left( \mathsf{S}_{0},\mathsf{L}_{0},%
\mathsf{H}_{0}\right) $. Markovianity here corresponds to the limit of
instantaneous feedback.

\begin{center}
%
%
%
%
%
%
%
%
%
\setlength{\unitlength}{.05cm}
\begin{picture}(120,80)
\label{pic5}

\thicklines
\put(30,40){\line(1,1){20}}

\thinlines
\put(10,50){\line(1,0){27}}
\put(10,50){\vector(1,0){10}}
\put(40,20){\line(0,1){27}}
\put(40,20){\vector(0,1){10}}
\put(40,53){\line(0,1){13}}
\put(40,53){\vector(0,1){10}}
\put(43,50){\line(1,0){77}}
\put(43,50){\vector(1,0){10}}
\put(120,20){\line(0,1){30}}
\put(40,20){\line(1,0){32}}
\put(120,20){\line(-1,0){32}}

\thicklines

\put(70,10){\line(0,1){20}}
\put(70,10){\line(1,0){20}}
\put(90,10){\line(0,1){20}}
\put(70,30){\line(1,0){20}}
\put(87,20){\circle{2}}
\put(73,20){\circle{2}}
\put(73,33){plant}

\put(10,55){$A_{\rm in}$} 
\put(40,74){$A_{\rm out}$}
\put(83,54){$B_{\rm in}$}

\put(33,14){$B_{\rm out}$}

\end{picture}
%

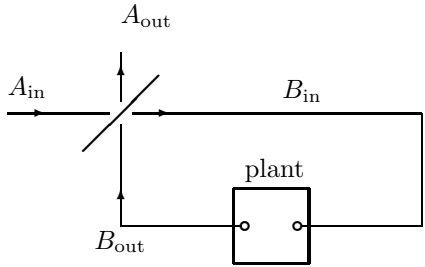
Figure 5: Feedback using a beam-splitter.
\end{center}

It is more convenient to view this as the network sketched below.

\begin{center}
\setlength{\unitlength}{.075cm}
\begin{picture}(60,50)
\label{pic6}
\thicklines
\put(25,10){$s_1$}
\put(25,20){$s_2$}
\put(25,40){$r_3$}
\put(31,10){$r_1$}
\put(31,20){$r_2$}
\put(31,40){$s_3$}
\put(20,5){\line(0,1){20}}
\put(20,5){\line(1,0){20}}
\put(40,5){\line(0,1){20}}
\put(20,25){\line(1,0){20}}
\put(20,35){\line(0,1){10}}
\put(20,35){\line(1,0){20}}
\put(40,35){\line(0,1){10}}
\put(20,45){\line(1,0){20}}
\thinlines
\put(10,20){\line(1,0){13}} 
\put(24,20){\circle{2}} 
\put(10,40){\vector(1,0){5}}
\put(10,10){\line(1,0){13}} 
\put(24,10){\circle{2}} 
\put(24,40){\circle{2}} 
\put(10,10){\vector(1,0){5}}
\put(37,20){\line(1,0){13}} 
\put(36,20){\circle{2}}
\put(37,10){\line(1,0){13}} 
\put(36,10){\circle{2}}
\put(36,40){\circle{2}}
\put(50,20){\vector(-1,0){5}}
\put(50,10){\vector(-1,0){5}}
\put(10,20){\line(0,1){20}}
\put(10,40){\line(1,0){13}}
\put(50,40){\line(-1,0){13}}
\put(50,20){\line(0,1){20}}
\put(2,5){$ A_{\rm out}$}
\put(55,5){$ A_{\rm in}$}
\put(43,43){{\it plant}}
\put(42,14){{\it beam splitter} }
\end{picture}
%

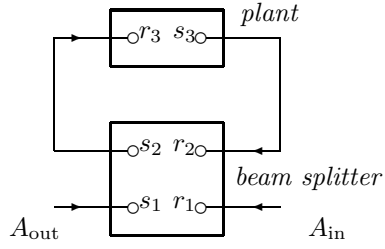
Figure 6: Network representation.
\end{center}

Here we have the pair of internal edges $\left( s_{2},r_{3}\right) $ and $%
\left( s_{3},r_{2}\right) $. The model matrix for the network is 
\begin{equation*}
\mathbf{V}=\left( 
\begin{array}{cccc}
-\frac{1}{2}\mathsf{L}_{0}^{\dag }\mathsf{L}_{0}-i\mathsf{H}_{0} & 0 & 0 & -%
\mathsf{L}_{0}^{\dag }\mathsf{S}_{0} \\ 
0 & T_{11} & T_{12} & 0 \\ 
0 & T_{21} & T_{22} & 0 \\ 
\mathsf{L}_{0} & 0 & 0 & \mathsf{S}_{0}
\end{array}
\right)
\end{equation*}
with respect to the labels $\left( 0,s_{1},s_{2},s_{3}\right) $ for the rows
and $\left( 0,r_{1},r_{2},r_{3}\right) $ for the columns. Here we have 
\begin{equation*}
\begin{tabular}{ll}
$\mathsf{S}_{\mathtt{ii}}=\left( 
\begin{array}{cc}
T_{22} & 0 \\ 
0 & \mathsf{S}_{0}
\end{array}
\right) ,$ & $\mathsf{S}_{\mathtt{ie}}=\left( 
\begin{array}{c}
T_{21} \\ 
0
\end{array}
\right) ,$ \\ 
$\mathsf{S}_{\mathtt{ei}}=\left( T_{12},0\right) ,$ & $\mathsf{S}_{\mathtt{ee%
}}=T_{11},$ \\ 
$\mathsf{L}_{\mathtt{i}}=\left( 
\begin{array}{c}
\mathsf{L}_{0} \\ 
0
\end{array}
\right) ,$ & $\mathsf{L}_{\mathtt{e}}=0,\quad \eta =\left( 
\begin{array}{cc}
0 & 1 \\ 
1 & 0
\end{array}
\right) .$%
\end{tabular}
\end{equation*}
Note that the adjacency matrix has row indices $\left( s_{2},s_{3}\right) $
and columns indices $\left( r_{2},r_{3}\right) $ labelling the internal
ports, and that the edges are the off diagonals $\left( s_{2},r_{3}\right) $
and $\left( s_{3},r_{3}\right) $. We may perform the two eliminations
simultaneously to obtain

\begin{eqnarray*}
\mathsf{S}_{\text{red}} &=&T_{11}+
\begin{array}{cc}
(T_{12} & 0) \\ 
& 
\end{array}
\left( 
\begin{array}{cc}
-T_{22} & 1 \\ 
1 & -\mathsf{S}_{0}
\end{array}
\right) ^{-1}\left( 
\begin{array}{c}
T_{21} \\ 
0
\end{array}
\right)  \\
&=&T_{11}+T_{12}\left( \mathsf{S}_{0}^{-1}-T_{22}\right) ^{-1}T_{21}, \\
\mathsf{L}_{\text{red}} &=&
\begin{array}{cc}
(T_{12} & 0) \\ 
& 
\end{array}
\left( 
\begin{array}{cc}
-T_{22} & 1 \\ 
1 & -\mathsf{S}_{0}
\end{array}
\right) ^{-1}\left( 
\begin{array}{c}
0 \\ 
\mathsf{L}_{0}
\end{array}
\right)  \\
&=&T_{12}\left( 1-\mathsf{S}_{0}T_{22}\right) ^{-1}\mathsf{L}_{0}, \\
\mathsf{H}_{\text{red}} &=&\mathsf{H}_{0}+\func{Im}
\begin{array}{cc}
(0 & \mathsf{L}_{0}^{\dag }) \\ 
& 
\end{array}
\left( 
\begin{array}{cc}
-T_{22} & 1 \\ 
1 & -\mathsf{S}_{0}
\end{array}
\right) ^{-1}\left( 
\begin{array}{c}
0 \\ 
\mathsf{L}_{0}
\end{array}
\right)  \\
&=&\mathsf{H}_{0}+\func{Im}\mathsf{L}_{0}^{\dag }\left( 1-\mathsf{S}%
_{0}T_{22}\right) ^{-1}\mathsf{L}_{0}\text{.}
\end{eqnarray*}

\subsection{The Redheffer Star Product}

An important feedback arrangement is shown in the figure below.

\begin{center}
\setlength{\unitlength}{.1cm}
\begin{picture}(60,65)
\label{pic7}
\thicklines
\put(20,10){\line(0,1){20}}
\put(20,10){\line(1,0){20}}
\put(40,10){\line(0,1){20}}
\put(20,30){\line(1,0){20}}
\put(20,40){\line(0,1){20}}
\put(20,40){\line(1,0){20}}
\put(40,40){\line(0,1){20}}
\put(20,60){\line(1,0){20}}
\thinlines
\put(24,14){\circle{2}}
\put(24,26){\circle{2}}
\put(36,14){\circle{2}}
\put(36,26){\circle{2}}
\put(24,44){\circle{2}}
\put(24,56){\circle{2}}
\put(36,44){\circle{2}}
\put(36,56){\circle{2}}
\put(23,14){\vector(-1,0){12}}
\put(49,14){\vector(-1,0){12}}
\put(11,56){\vector(1,0){12}}
\put(37,56){\vector(1,0){12}}
\put(23,26){\line(-1,0){8}}
\put(15,26){\vector(0,1){18}}
\put(23,44){\line(-1,0){8}}
\put(37,26){\line(1,0){8}}
\put(45,44){\vector(0,-1){18}}
\put(37,44){\line(1,0){8}}
\put(28,62){$A$}
\put(0,12){${\bf b}^{\rm out}_4$}
\put(-5,35){$ {\bf b}^{\rm out}_3  ={\bf b}^{\rm in}_2$}
\put(0,56){${\bf b}^{\rm in}_1$}
\put(28,4){$B$}
\put(50,12){${\bf b}^{\rm in}_4$}
\put(50,35){${\bf b}^{\rm out}_2 = {\bf b}^{\rm in}_3$}
\put(50,56){${\bf b}^{\rm out}_1$}
\put(26,14){$s_4$}
\put(26,26){$s_3$}
\put(26,44){$r_2$}
\put(26,56){$r_1$}
\put(31,14){$r_4$}
\put(31,26){$r_3$}
\put(31,44){$s_2$}
\put(31,56){$s_1$}
\end{picture}%
%

Figure 7 Composite System
\end{center}

We shall now derive this system taking component $A$ to be described $\left( 
\begin{array}{cc}
\mathsf{S}_{11}^{A} & \mathsf{S}_{12}^{A} \\ 
\mathsf{S}_{21}^{A} & \mathsf{S}_{22}^{A}
\end{array}
\right) ,$ $\left( 
\begin{array}{c}
\mathsf{L}_{1}^{A} \\ 
\mathsf{L}_{2}^{A}
\end{array}
\right) ,$ $\mathsf{H}_{A}$ and $B$ by $\left( 
\begin{array}{cc}
\mathsf{S}_{33}^{B} & \mathsf{S}_{34}^{B} \\ 
\mathsf{S}_{43}^{B} & \mathsf{S}_{44}^{B}
\end{array}
\right) ,$ $\left( 
\begin{array}{c}
\mathsf{L}_{3}^{B} \\ 
\mathsf{L}_{4}^{B}
\end{array}
\right) ,$ $\mathsf{H}_{B}$. The operators of systems $A$ are assumed to
commute with those of $B$. We have two internal channels to eliminate which
we can do in sequence, or simultaneously. We shall do the latter. here we
have 
\begin{eqnarray*}
\mathsf{S}_{\mathtt{ee}} &=&\left( 
\begin{array}{cc}
\mathsf{S}_{11}^{A} & 0 \\ 
0 & \mathsf{S}_{44}
\end{array}
\right) ,\mathsf{S}_{\mathtt{ei}}=\left( 
\begin{array}{cc}
\mathsf{S}_{12}^{A} & 0 \\ 
0 & \mathsf{S}_{43}^{B}
\end{array}
\right)  \\
\mathsf{S}_{\mathtt{ie}} &=&\left( 
\begin{array}{cc}
\mathsf{S}_{21}^{A} & 0 \\ 
0 & \mathsf{S}_{34}^{B}
\end{array}
\right) ,\mathsf{S}_{\mathtt{ii}}=\left( 
\begin{array}{cc}
\mathsf{S}_{22}^{A} & 0 \\ 
0 & \mathsf{S}_{33}^{B}
\end{array}
\right) 
\end{eqnarray*}
and 
\begin{equation*}
\mathsf{L}_{\mathtt{e}}=\left( 
\begin{array}{c}
\mathsf{L}_{1}^{A} \\ 
\mathsf{L}_{4}^{B}
\end{array}
\right) ,\quad \mathsf{L}_{\mathtt{i}}=\left( 
\begin{array}{c}
\mathsf{L}_{2}^{A} \\ 
\mathsf{L}_{3}^{B}
\end{array}
\right) ,\quad \eta =\left( 
\begin{array}{cc}
0 & 1 \\ 
1 & 0
\end{array}
\right) .
\end{equation*}
The reduced operators are therefore 
\begin{eqnarray*}
\mathsf{S}_{\star } &=&\left( 
\begin{array}{cc}
\mathsf{S}_{11}^{A} & 0 \\ 
0 & \mathsf{S}_{44}^{B}
\end{array}
\right) +\left( 
\begin{array}{cc}
\mathsf{S}_{12}^{A} & 0 \\ 
0 & \mathsf{S}_{43}^{B}
\end{array}
\right) \left( 
\begin{array}{cc}
-\mathsf{S}_{22}^{A} & 1 \\ 
1 & -\mathsf{S}_{33}^{B}
\end{array}
\right) ^{-1}\left( 
\begin{array}{cc}
\mathsf{S}_{21}^{A} & 0 \\ 
0 & \mathsf{S}_{34}^{B}
\end{array}
\right)  \\
&=&\left( 
\begin{array}{cc}
\mathsf{S}_{11}^{A}+\mathsf{S}_{12}^{A}\mathsf{S}_{33}^{B}\left( 1-\mathsf{S}%
_{22}^{A}\mathsf{S}_{33}^{B}\right) ^{-1}\mathsf{S}_{21}^{A} & \mathsf{S}%
_{12}^{A}\left( 1-\mathsf{S}_{22}^{A}\mathsf{S}_{33}^{B}\right) ^{-1}\mathsf{%
S}_{34}^{B} \\ 
\mathsf{S}_{43}^{B}\left( 1-\mathsf{S}_{22}^{A}\mathsf{S}_{33}^{B}\right)
^{-1}\mathsf{S}_{21}^{A} & \mathsf{S}_{44}+\mathsf{S}_{43}^{B}\left( 1-%
\mathsf{S}_{22}^{A}\mathsf{S}_{33}^{B}\right) ^{-1}\mathsf{S}_{22}^{A}%
\mathsf{S}_{34}^{B}
\end{array}
\right) ,
\end{eqnarray*}
\begin{eqnarray*}
\mathsf{L}_{\star } &=&\left( 
\begin{array}{c}
\mathsf{L}_{1}^{A} \\ 
\mathsf{L}_{4}^{B}
\end{array}
\right) +\left( 
\begin{array}{cc}
\mathsf{S}_{12}^{A} & 0 \\ 
0 & \mathsf{S}_{43}^{B}
\end{array}
\right) \left( 
\begin{array}{cc}
-\mathsf{S}_{22}^{A} & 1 \\ 
1 & -\mathsf{S}_{33}^{B}
\end{array}
\right) ^{-1}\left( 
\begin{array}{c}
\mathsf{L}_{2}^{A} \\ 
\mathsf{L}_{3}^{B}
\end{array}
\right)  \\
&=&\left( 
\begin{array}{c}
\mathsf{L}_{1}^{A}+\mathsf{S}_{12}^{A}\mathsf{S}_{33}^{B}\left( 1-\mathsf{S}%
_{22}^{A}\mathsf{S}_{33}^{B}\right) ^{-1}\mathsf{L}_{2}^{A}+\mathsf{S}%
_{12}^{A}\left( 1-\mathsf{S}_{22}^{A}\mathsf{S}_{33}^{B}\right) ^{-1}\mathsf{%
L}_{3}^{B} \\ 
\mathsf{L}_{4}^{B}+\mathsf{S}_{43}^{B}\left( 1-\mathsf{S}_{22}^{A}\mathsf{S}%
_{33}^{B}\right) ^{-1}\mathsf{L}_{2}^{A}+\mathsf{S}_{43}^{B}\mathsf{S}%
_{22}^{A}\left( 1-\mathsf{S}_{22}^{A}\mathsf{S}_{33}^{B}\right) ^{-1}\mathsf{%
L}_{3}^{B}
\end{array}
\right) ,
\end{eqnarray*}
\begin{equation*}
\begin{array}{l}
\mathsf{H}_{\star }=\mathsf{H}_{A}+\mathsf{H}_{B}+\func{Im}\left\{ \mathsf{L}%
_{3}^{B\dag }\left( 1-\mathsf{S}_{33}^{B}\mathsf{S}_{22}^{A}\right) ^{-1}%
\mathsf{L}_{3}^{B}+\mathsf{L}_{3}^{B\dag }\left( 1-\mathsf{S}_{33}^{B}%
\mathsf{S}_{22}^{A}\right) ^{-1}\mathsf{S}_{33}^{B}\mathsf{L}_{2}^{A}\right. 
\\ 
+\mathsf{L}_{2}^{A\dag }\left( 1-\mathsf{S}_{22}^{A}\mathsf{S}%
_{33}^{B}\right) ^{-1}\mathsf{S}_{22}^{A}\mathsf{L}_{3}^{B}+\mathsf{L}%
_{2}^{A\dag }\left( 1-\mathsf{S}_{22}^{A}\mathsf{S}_{33}^{B}\right) ^{-1}%
\mathsf{L}_{2}^{A} \\ 
+\mathsf{L}_{1}^{A\dag }\mathsf{S}_{12}^{A}\left( 1-\mathsf{S}_{33}^{B}%
\mathsf{S}_{22}^{A}\right) ^{-1}\mathsf{L}_{3}^{B}+\mathsf{L}_{1}^{A\dag }%
\mathsf{S}_{12}^{A}\left( 1-\mathsf{S}_{33}^{B}\mathsf{S}_{22}^{A}\right)
^{-1}\mathsf{S}_{33}^{B}\mathsf{L}_{2}^{A} \\ 
\left. +\mathsf{L}_{4}^{B\dag }\mathsf{S}_{43}^{B}\left( 1-\mathsf{S}%
_{22}^{A}\mathsf{S}_{33}^{B}\right) ^{-1}\mathsf{S}_{22}^{A}\mathsf{L}%
_{3}^{B}+\mathsf{L}_{4}^{B\dag }\mathsf{S}_{43}^{B}\left( 1-\mathsf{S}%
_{22}^{A}\mathsf{S}_{33}^{B}\right) ^{-1}\mathsf{L}_{2}^{A}\right\} .
\end{array}
\end{equation*}

\section{Topological Rules for Quantum Networks}

We wish to state algebraic rules for constructing the operators $\mathsf{S}_{%
\text{net}},\mathsf{L}_{\text{net}},\mathsf{H}_{\text{net}}$ covering all
the examples of feedback networks studied so far. For a given network, we
denote by $\mathcal{P}_{\text{out}}$ the set of output ports and $\mathcal{P}%
_{\text{out}}^{\text{ext}}$ the subset of output ports having external
output. With each $i\in \mathcal{P}_{\text{out}}$ there is an associated 
\textit{port operator} \textsf{$L$}$_{i}$. Similarly we have $\mathcal{P}_{%
\text{in}}$ and $\mathcal{P}_{\text{in}}^{\text{ext}}$ for the inputs.

For $\gamma =\left( i,j\right) $ an ordered pair of ports, we set 
\begin{equation*}
\mathsf{S}_{\gamma }=\left\{ 
\begin{tabular}{ll}
1, & if $j\in \mathcal{P}_{\text{out}}$ and $i\in \mathcal{P}_{\text{in}}$
and there is a feedback connection \\ 
& from $j$ to $i$; \\ 
$\mathsf{S}_{ij},$ & if $i\in \mathcal{P}_{\text{out}}$ and $j\in \mathcal{P}%
_{\text{in}}$ and both $i$ and $j$ are in the same plant \\ 
& and where $\mathsf{S}_{ij}$ is the scattering component from $j$ to $i$;
\\ 
0, & otherwise.
\end{tabular}
\right.
\end{equation*}
More generally, if $\gamma $ is an ordered sequence (paths) of ports,
alternating between input and output ports, we define \textsf{$S$}$_{\gamma
} $ inductively. If $\gamma $ is the concatenation of $\gamma _{1}$ followed
by $\gamma _{2}$ then 
\begin{equation*}
\mathsf{S}_{\gamma }=\mathsf{S}_{\gamma _{2}}\mathsf{S}_{\gamma _{1}}.
\end{equation*}
Finally, we set $\Gamma \left( i,j\right) $ to be the set of all paths going
from $j$ to $i$. Then for $i\in \mathcal{P}_{\text{out}}^{\text{ext}}$, $%
j\in \mathcal{P}_{\text{in}}^{\text{ext}}$%
\begin{equation}
\begin{array}{c}
\\ 
\multicolumn{1}{l}{\left( \mathsf{S}_{\text{net}}\right) _{ij}=\sum_{\gamma
\in \Gamma \left( i,j\right) }\mathsf{S}_{\gamma };} \\ 
\\ 
\multicolumn{1}{l}{\left( \mathsf{L}_{\text{net}}\right) _{i}=\sum_{k\in 
\mathcal{P}_{\text{out}}}\sum_{\gamma \in \Gamma \left( i,k\right) }\mathsf{S%
}_{\gamma }\mathsf{L}_{k};} \\ 
\\ 
\multicolumn{1}{l}{\mathsf{H}_{\text{net}}=\mathsf{H}_{0}+\sum_{l,k\in 
\mathcal{P}_{\text{out}}}\sum_{\gamma \in \Gamma \left( l,k\right) }\mathsf{L%
}_{l}^{\dag }\mathsf{S}_{\gamma }\mathsf{L}_{k},} \\ 
\text{ }
\end{array}
\end{equation}
where $\mathsf{H}_{0}$ is the sum of the individual component systems
Hamiltonians.

\begin{acknowledgement}
The authors would like to thank M. Yanagisawa for several\ useful comments
in writing this paper, and to L. Bouten for suggesting the Trotter-Kato
convergence technique in the main theorem. They are also grateful to the
referee for bringing several important technical points to their attention.
\end{acknowledgement}


\bigskip

\begin{thebibliography}{99}
\bibitem{AH}  L. Accardi, R.L. Hudson: \textit{Quantum stochastic flows and
non-abelian cohomology}\ Quantum Probability V, Lecture Notes in Mathematics 
\textbf{1442}, 54-69 (1990)

\bibitem{Belavkin79}  V. P. Belavkin: Optimization of Quantum Observation
and Control, In Proc. of 9th IFIP Conf on Optimizat Techn. Notes in Control
and Inform Sci \textbf{1}, Springer-Verlag, Warszawa (1979)

\bibitem{Belavkin83}  V. P. Belavkin: Theory of the Control of Observable
Quantum Systems. Automatica and Remote Control \textbf{44} (2) 178--188
(1983).

\bibitem{Belavkin CCR}  V.P. Belavkin. Quantum continual measurements and a
posteriori collapse on CCR. Commun. Math. Phys., \textbf{146}:611635, (1992)

\bibitem{Belavkin98}  V. P. Belavkin: On Quantum Ito Algebras and Their
Decompositions. Letters in Mathematical Physics \textbf{45}, 131-145 (1998)

\bibitem{Bouten vHandel James}  L. Bouten, R. Van Handel, and M.R. James. An
introduction to quantum filtering. math.OC/06011741 (2006)

\bibitem{Carmichael}  H.J. Carmichael. Quantum trajectory theory for
cascaded open systems. Phys. Rev. Lett., \textbf{70}(15):22732276, (1993)

\bibitem{Caves}  C.M. Caves, Quantum limits on noise in linear amplifiers,
Phys. Rev. D., \textbf{26}, 1817, (1982)

\bibitem{Cheb98}  A.M. Chebotarev, Quantum stochastic differential equation
is unitarily equivalent to a symmetric boundary problem in Fock space, Inf.
Dim. Anal. Quantum Prob., \textbf{1}, 175-199, (1998)

\bibitem{Davies}  E.B. Davies, One-parameter Semigroups, Academic Press Inc,
London (1980)

\bibitem{Gardiner}  C.W. Gardiner. Driving a quantum system with the output
field from another driven quantum system. Phys. Rev. Lett., \textbf{70}%
(15):22692272, 1993.

\bibitem{Gardiner Collett}  C.W. Gardiner and M.J. Collett. Input and output
in damped quantum systems: Quantum stochastic differential equations and the
master equation. Phys. Rev. A, \textbf{31}(6):37613774, (1985)

\bibitem{GQCLT}  J. Gough, Quantum flows as Markovian limit of emission,
absorption and scattering interactions, Commun. Math. Phys. \textbf{254},
no. 2, 489-512 (2005)

\bibitem{GBS}  J. Gough, V.P Belavkin, O.G. Smolyanov,
Hamilton-Jacobi-Bellman equations for Quantum Filtering and Control, J. Opt.
B: Quantum Semiclass. Opt. \textbf{S237-244}, Special issue on quantum
control, (2005)

\bibitem{GJ}  J. Gough, M.R. James, The series product and its application
to feedforward and feedback networks, arXiv:07070048(v1) [quant-ph], to appear IEEE Trans.
Automatic Control

\bibitem{GreenLimebeer}  M. Green and D. J. N. Limebeer, Linear Robust
Control, Prentice Hall, New Jersey 07632, U.S.A. (1995)

\bibitem{Gregoratti}  M. Gregoratti, The Hamiltonian operator associated to
some quantum stochastic differential equations, Commun. Math. Phys., \textbf{%
222}, 181-200, (2001)

\bibitem{HP}  R.L. Hudson and K.R. Parthasarathy. Quantum Ito's formula and
stochastic evolutions. Commun. Math. Phys., 93:301323, (1984)

\bibitem{S Lloyd}  S. Lloyd. Coherent quantum feedback. Phys. Rev. A,
62:022108, (2000)

\bibitem{Warszawski Wiseman Mabuchi}  P. Warszawski, H.M. Wiseman, and H.
Mabuchi. Quantum trajectories for realistic detection. Phys. Rev. A,
65:023802, (2002)

\bibitem{Wiseman Feedback}  H. Wiseman. Quantum theory of continuous
feedback. Phys. Rev. A, 49(3):21332150, (1994)

\bibitem{YK1}  M. Yanagisawa and H. Kimura. Transfer function approach to
quantum control-part I: Dynamics of quantum feedback systems. IEEE Trans.
Automatic Control, (48):21072120, 12 (2003)

\bibitem{YK2}  M. Yanagisawa and H. Kimura. Transfer function approach to
quantum control-part II: Control concepts and applications. IEEE Trans.
Automatic Control, (48):21212132, 12 (2003)

\bibitem{Young}  N. Young, An Introduction to Hilbert Space, Cambridge
Mathematical Textbooks, (1988)

\bibitem{Yurke Denker}  B. Yurke and J.S. Denker. Quantum network theory.
Phys. Rev. A, \textbf{29}(3):14191437, (1984)

\bibitem{ZDG96}  K.~Zhou and J.~Doyle and K.~Glover. Robust and Optimal
Control. Prentice Hall, NJ, 1996.
\end{thebibliography}
\end{document}